\global\def\draftcontrol{0}

\def\versionno{$\beta=0$}

\catcode`\@=11

\expandafter\ifx\csname draftcontrol\endcsname\relax\global\def\draftcontrol{0} 
\fi 

{\count255=\time\divide\count255 by 60 
\xdef\hourmin{\number\count255} 
\multiply\count255 by-60\advance\count255 by\time 
\xdef\hourmin{\hourmin:\ifnum\count255<10 0\fi\the\count255}} 
\def\draftdate{\number\month/\number\day/\number\year\ \ \ \hourmin } 

\newcommand\makepapertitle{\par

\begingroup 
 \renewcommand\thefootnote{\@fnsymbol\c@footnote}%
 \def\@makefnmark{\rlap{\@textsuperscript{\normalfont\@thefnmark}}}%
 \long\def\@makefntext##1{\parindent 1em\noindent 
         \hb@xt@1.8em{%
             \hss\@textsuperscript{\normalfont\@thefnmark}}##1}%
  \newpage 
  \global\@topnum\z@   
  \@makepapertitle 
  \thispagestyle{empty}\@thanks 
\endgroup 
\setcounter{footnote}{0}%
\global\let\thanks\relax 
\global\let\makepapertitle\relax 
\global\let\@makepapertitle\relax 
\global\let\@thanks\@empty 
\global\let\@author\@empty 
\global\let\@date\@empty 
\global\let\@title\@empty 
\global\let\title\relax 
\global\let\author\relax 
\global\let\date\relax 
\global\let\and\relax 
\def\version{\let\version\@version\@gobble} 
} 
\def\@makepapertitle{%
\newpage 
\ifnum\draftcontrol=1 {} 
\version\versionno 
\vskip 5.5em%
\else 
\hfill\hbox to 3cm {\parbox{4.5cm}{\@pubnum}\hss}%
\vskip 6.5em%
\fi 
\begin{center}%
\let \footnote \thanks 
   {\hskip -0\textwidth \hbox to 1\textwidth%
     {\centerline{\Large\bf{\noindent\@title}}}}%
  \vskip 2em%
  {\normalsize
    \lineskip .5em%
    \begin{tabular}[t]{c}%
      \@author 
    \end{tabular}\par}%
  \vskip 1.5em%
  {\@bstract}%
  \end{center}%
  \vfill
  \@date%
  \vskip 1.5em%
\par 
} 

\gdef\@pubnum{} 
\def\pubnum#1{%
\gdef\@pubnum{#1}} 

\gdef\@bstract{} 
\def\Abstract#1{%
\gdef\@bstract{%
\parbox{\textwidth-0pc}{%
\centerline{\bf Abstract}\penalty1000 
\noindent
\renewcommand\baselinestretch{1.0} 
{#1}}} 
} 

\gdef\@email{}
\def\email#1{%
\gdef\@email{%
Email: {\tt #1}}
}

\def\ps@paper{\let\@mkboth\@gobbletwo%
  \ifnum\draftcontrol=1 
     \def\@oddfoot{\hbox to \textwidth{\tiny \versionno \hfil\tiny\draftdate}%
     \hskip -\textwidth \hbox to \textwidth{\hfil\rm\thepage\hfil}}%
  \else\def\@oddfoot{\hbox to \textwidth{\hfil\rm\thepage\hfil}} 
  \fi 
  \let\@evenfoot\@oddfoot 
} 

\def\body{\clearpage 
       \pagestyle{paper} 
     } 
\newenvironment{acknowledgments}{%
\vskip 3.25ex 
\addcontentsline{toc}{section}{Acknowledgments}
\noindent {\bf Acknowledgments} 
} 

\def\@version#1{\ifnum\draftcontrol=1 
\typeout{}\typeout{#1}\typeout{} 
\vskip3mm\centerline{\hbox{\fbox{\normalsize{\tt DRAFT -- #1 -- } 
                {\draftdate}}}}\vskip3mm 
\fi} 
\let\version\@version 
\long\def\eqlabel#1{\ifnum\draftcontrol=1 
                 \tag@false  
                 \tag*{(\theequation) \hbox to -0.2cm{\hspace{0cm}\small{#1}\hss}} 
                 \refstepcounter{equation}  
                 \edef\@currentlabel{\theequation} 
                 \ltx@label{#1}          
                 \else 
                 \label{#1} 
                 \fi 
                 } 
\let\st@bibitem\@bibitem 
\let\st@lbibitem\@lbibitem 
\ifnum\draftcontrol=1 
\def\@bibitem#1{%
 \st@bibitem{#1}\a@@label{#1}\ignorespaces} 
\def\@lbibitem[#1]#2{%
 \st@lbibitem[#1]{#2}\a@@label{#2}\ignorespaces} 
\def\a@@label#1{%
 \gdef\a@lab{\smash{\normalfont\small#1}} 
 \ifvmode 
   \if@inlabel 
     \global\setbox\@labels\hbox{%
       \llap{\a@lab\let\a@lab\relax 
             \kern\@totalleftmargin\kern\marginparsep}%
       \box\@labels}%
   \fi 
 \fi} 
\fi 

\documentclass[12pt,letterpaper]{article} 

\usepackage{amsmath,bm,amsfonts,amssymb,array,calc,amsthm,rotating}
\usepackage{epsfig,psfrag} 
\usepackage{rotating}
\usepackage{amscd}
\usepackage{graphicx}
\usepackage{color}
\usepackage[colorlinks=false]{hyperref}

\renewcommand\baselinestretch{1.25} 
\renewcommand\section{\@startsection {section}{1}{\z@}%
                                {-3.5ex \@plus -1ex \@minus -.2ex}%
                                {2.3ex \@plus.2ex}%
                                {\normalfont\large\bfseries}} 
\renewcommand\subsection{\@startsection{subsection}{2}{\z@}%
                                {-3.25ex\@plus -1ex \@minus -.2ex}%
                                {1.5ex \@plus .2ex}%
                                {\normalfont\normalsize\bfseries}} 
\renewcommand\subsubsection{\@startsection{subsubsection}{3}{\z@}%
                                {-3.25ex\@plus -1ex \@minus -.2ex}%
                                {1.5ex \@plus .2ex}%
                                {\normalfont\normalsize\it}} 
\renewcommand\paragraph{\@startsection{paragraph}{4}{\z@}%
                                {-1.75ex\@plus -1ex \@minus -.2ex}%
                                {1ex \@plus .2ex}%
                                {\normalfont\normalsize\bf}} 
\renewcommand\subparagraph{\@startsection{subparagraph}{5}{\z@}%
                                {-1.25ex\@plus -0ex \@minus -.2ex}%
                                {-2ex \@plus .2ex}%
                                {\normalfont\normalsize\it}}

\numberwithin{equation}{section}

\long\def\@makecaption#1#2{%
\vskip\abovecaptionskip
\sbox\@tempboxa{{\bf #1:} #2}%
\ifdim \wd\@tempboxa >\hsize
 {\small\bf #1:} {\small #2}\par
\else
 \global \@minipagefalse
 \hb@xt@\hsize{\hfil\box\@tempboxa\hfil}%
\fi
\vskip\belowcaptionskip}

\renewcommand*\l@section[2]{%
\ifnum \c@tocdepth >\z@
 \addpenalty\@secpenalty
 \addvspace{.5em \@plus\p@}%
 \setlength\@tempdima{1.5em}%
 \begingroup
   \parindent \z@ \rightskip \@pnumwidth
   \parfillskip -\@pnumwidth
   \leavevmode \bfseries
   \advance\leftskip\@tempdima
   \hskip -\leftskip
   #1\nobreak\hfil \nobreak\hb@xt@\@pnumwidth{\hss #2}\par
 \endgroup
\fi}
\renewcommand*\l@subsection{\addvspace{.0em \@plus\p@}\@dottedtocline{2}{1.5em}{2.3em}}
\renewcommand*\l@subsubsection{\addvspace{-.2em \@plus\p@}\@dottedtocline{3}{3.8em}{3.2em}}

\def\hepth#1{\href{http://xxx.arxiv.org/abs/hep-th/#1}{{arXiv:hep-th/#1}}}
\def\mathph#1{\href{http://xxx.arxiv.org/abs/math-ph/#1}{{arXiv:math-ph/#1}}}

\def\mathag#1{\href{http://xxx.arxiv.org/abs/math.AG/#1}{{arXiv:math.ag/#1}}}

\def\solvint#1{\href{http://xxx.arxiv.org/abs/solv-int/#1}{{arXiv:solv-int/#1}}}
\def\arxiv#1#2{\href{http://xxx.arxiv.org/abs/#1}{{arXiv:#1 [#2]}}}

\definecolor{refcol}{rgb}{0.2,0.2,0.8}
\definecolor{eqcol}{rgb}{.6,0,0}
\definecolor{purple}{cmyk}{0,1,0,0}

\gdef\@citecolor{refcol}
\gdef\@linkcolor{eqcol}
\def\colorlinkspurple{\gdef\@urlcolor{purple}}
\def\colorlinksblue{\gdef\@urlcolor{blue}}
\def\colorlinksred{\gdef\@urlcolor{red}}

\def\ie{{\it i.e.}}

\def\cf{{\it cf.}}

\def\revise#1       {\raisebox{-0em}{\rule{3pt}{1em}}%
                  \marginpar{\raisebox{.5em}{\vrule width3pt\ 
                  \vrule width0pt height 0pt depth0.5em 
                  \hbox to 0cm{\hspace{0cm}{%
                  \parbox[t]{4em}{\raggedright\footnotesize{#1}}}\hss}}}}

\def\del          {\partial} 

\def\ii           {{\it i}}

\def\sqr#1#2{{\vcenter{\vbox{\hrule height.#2pt   
\hbox{\vrule width.#2pt height#1pt \kern#1pt 
\vrule width.#2pt}\hrule height.#2pt}}}}

\newcommand{\beq}{\begin{equation}}
\newcommand{\eq}{\end{equation}}
\newcommand{\req}[1]{(\ref{#1})}
\renewcommand{\ie}{{\it i.e.}}

\renewcommand{\t}{\tilde}

\newcommand{\Fcal}{\mathcal F}

\newcommand{\Wcal}{\mathcal W}
\newcommand{\Ocal}{\mathcal O}

\newcommand{\Ccal}{\mathcal C}
\newcommand{\Dcal}{\mathcal D}

\renewcommand{\P}{\mathbb P}

\newcommand{\R}{\mathbb R}
\newcommand{\T}{\mathbb T}
\newcommand{\C}{\mathbb C}

\newcommand{\ep}{\epsilon}

\newcommand{\bea}{\begin{eqnarray}}
\newcommand{\eea}{\end{eqnarray}}
\newcommand{\bee}{\begin{eqnarray*}}
\newcommand{\eee}{\end{eqnarray*}}
\newcommand{\be}{\begin{equation}}
\newcommand{\ee}{\end{equation}}
\newcommand{\bem}{\begin{pmatrix}}
\newcommand{\eem}{\end{pmatrix}}

\def\g{\gamma}
\def\e{\epsilon}
\def\b{\beta}
   
\def\pa{\partial}       
\def\inf{\infty}       
\def\til{\tilde}
\def\Tr{{\rm Tr}}
\def\p{\pi}

\catcode`\@=12 

\begin{document} 


\title{Quantum Geometry of Refined Topological Strings}

\pubnum{
UCB-PTH-11/03
}
\date{April 2011}

\author{
Mina Aganagic$^{a, b}$, Miranda C. N. Cheng$^{c,d,e}$, Robbert Dijkgraaf$^{f}$,    \\[0.2cm]
Daniel Krefl$^{b}$ and Cumrun Vafa$^{d}$ \\[0.4cm]
\it $^{a}$ Department of Mathematics, University of California, Berkeley, USA\\
\it $^{b}$ Center for Theoretical Physics, University of California, Berkeley, USA\\
\it $^{c}$ Department of Mathematics, Harvard University, Cambridge, USA\\
\it $^{d}$ Department of Physics, Harvard University, Cambridge, USA\\
\it $^{e}$ School of Natural Sciences, Institute for Advanced Study, Princeton, USA\\
\it $^{f}$ Institute for Theoretical Physics, University of Amsterdam, The Netherlands
}

\Abstract{We consider branes in refined topological strings.  We argue
that their wave-functions satisfy a Schr\"odinger equation depending on multiple times and prove this in the case where the
topological string has a dual matrix model description. Furthermore, in the limit where one of the equivariant rotations approaches zero,
the brane partition function satisfies a time-independent Schr\"odinger equation.  We use
this observation, as well as the back reaction of the brane on the closed string geometry, to 
offer an explanation of the connection between integrable systems
and ${\cal N}=2$ gauge systems in four dimensions observed by Nekrasov and Shatashvili.
}

\makepapertitle

\body

\version\versionno

\vskip 1em

\tableofcontents

\newpage

\section{Introduction}
\label{Introduction}

The study of topological strings on Calabi-Yau threefolds has led to important insights in various aspects of
string theory and supersymmetric gauge theories.  For instance, the properties of topological
strings on compact Calabi-Yau geometries has impacted our understanding of BPS states of four-dimensional
charged black holes.  On the other hand, the study of topological strings in the background of non-compact `local'  Calabi-Yau geometries
has enriched our understanding of ${\cal N}=2$ and ${\cal N}=1$ supersymmetric gauge theories in four dimensions.

More recently, inspired by the work of Nekrasov on instanton partition functions of ${\cal N}=2$
gauge theories in four and five dimensions \cite{nekrasov}, it became clear that there should exist a refinement of topological
strings for the case of non-compact Calabi-Yau threefolds.  
This is because these gauge theories can also be engineered using non-compact, toric Calabi-Yau threefolds, for which the topological
string partition function is a special limit of the instanton partition function, \ie, 
$$Z_{\text{\it top}}(g_s)=Z_{\text{\it{Nek}}}(\epsilon_1,\epsilon_2)\lvert_{-\ep_1=\ep_2=g_s}\,,$$
where $\epsilon_\alpha$ denote the two equivariant rotations of space-time (taken to be ${\C}^2$), and $g_s$ is the coupling constant of topological strings.
This connection implies that there should be a refinement of the topological string partition function, where the string coupling constant is split into two
independent parameters.

Despite of the above hint, a world-sheet description of such a refined topological string is still lacking. Nevertheless, the above observation does motivate a definition for the refined A-model topological string in terms of its lift to M-theory and the degeneracy
of BPS states in five dimensions.  In particular, the standard topological string partition function captures
the BPS degeneracy of M2-branes wrapped over 2-cycles.  Furthermore, the topological string
coupling constant is related to the chemical potential for the $SU(2)_L\subset SU(2)_L\times SU(2)_R=SO(4)$ rotation group
in five dimensions.  The refinement of the topological string captures in addition the $SU(2)_R$ rotation
quantum numbers of the BPS states, and thus the full rotational quantum numbers of the BPS states.

Independently, in the context of B-model topological strings that are dual to matrix models, an alternative definition was proposed in \cite{DV09},
where the 
refinement involves the so-called `$\beta$-ensemble' of matrix models.  This is just the ordinary
matrix model, but now with the power of the Vandermonde determinant of the eigenvalues raised to $2\beta$. The parameter $\beta$ is related to the equivariant rotation parameters via $\beta=-\ep_1/\ep_2$ and equals to 1 in the case of the usual topological string.

One natural question is how to actually compute the partition function of the refined topological
string.  There are various possibilities.  For the A-model, it has been proposed that
the refined topological vertex, which is the one-parameter refinement of the standard topological vertex,
computes the corresponding degeneracies \cite{ikv}.  Another approach, in the context of B-model, is to use the standard
matrix model techniques but adapted to the $\beta$-ensemble \cite{DV09}.  Finally, yet another approach is to use
the fact that the refined topological string partition function is still expected to be a wave-function,
and thus satisfies the holomorphic anomaly equation.  In this way the computation of the partition function
reduces to fixing the holomorphic ambiguities \cite{kw1,kw2,hk10}.  Even though these different approaches have not been
proven to be equivalent, so far they give results which agree with each other and with the results of Nekrasov in the cases that have been checked.

On the other hand, an interesting observation concerning the instanton partition function was made by Nekrasov
and Shatashvili (NS) \cite{nesh}. They noticed that in the limit where one of the equivariant parameters is sent to zero while the other is kept fixed, for instance $\epsilon_1\rightarrow 0$ with $\epsilon_2 = \hbar$ fixed, a connection with certain quantum integrable systems emerges in the following way: The integrally stationary equation of the free energy
\beq\eqlabel{NSintcond}
\exp (\partial_{a_I}{\cal W}(\vec a;\hbar))=1\,,
\eq
where $${\cal W}(\vec a;\hbar)=\lim_{\ep_1\to0} \epsilon_1{\cal F}(\vec a; \epsilon_1,\epsilon_2=\hbar)\quad,\quad {\cal F} = \log Z_{\text{\it Nek}}$$ and $\vec a$ denotes the collection of all Coulomb parameters $\{a_{I}\}$ of the theory,
gives the Bethe ansatz equation for the corresponding integrable system.  In particular the energy eigenvalues of the integrable
system are completely characterized by the $a_I$ satisfying these equations.  Even though this is a striking statement,
and was checked in many cases, an explanation of it was not offered.  One aim of this paper is to shed light on this
relation.

Our approach is to study branes in the context of the refined topological string.  
It has been long known that the branes probe geometry in a quantum mechanical way. In particular, in the context of the standard, unrefined topological string, it was argued in  \cite{adkmv} that for the  B-model on a local Calabi-Yau  given by
$$uv+H(x,p)=0\,,$$
the wave-function $\Psi(x)$ of a brane whose position is labeled by a point $x$ on the Riemann
surface $H(x,p)=0$ classically, satisfies an operator equation 
\beq\eqlabel{Hpsisec1}
H\Psi(x) =0\,,
\eq
at quantum level, with $H:=H(x,g_s \partial_x)$.  In other words, one is studying a quantum mechanical
problem given by the Hamiltonian $H$.  Clearly, this is the right context to elucidate the connection to the results of NS integrability,
since the B-model mirror to a geometrically engineered gauge theory is indeed a geometry of the above form
and the corresponding integrable model is specified by $H$ defined above. 
However, as pointed out in \cite{adkmv}, when there are cycles in the Calabi-Yau geometry corresponding to magnetic charges there are further $g_s$ corrections to $H$ itself. Therefore, in general the above operator relation only holds up to order $g_s$ corrections.

In this paper we consider brane partition functions in the context of refined topological strings. 
In this case there are two types of branes, corresponding in the M-theory language to M5 branes wrapping a Lagrangian cycle in the Calabi-Yau and a two-plane $z_1$ or $z_2$ in ${\mathbb R}^4\simeq {\mathbb C}^2$ associated with the $\ep_1$  and $\ep_2$ action respectively.
In the B-model context, we show that for a $\beta$-deformed matrix model with a polynomial potential, the equation
\req{Hpsisec1} is generalized to an equation which is exact and of the form of a multi-time dependent version
of the Schr\"odinger equation, \ie, 
$$H\Psi =\ep_1\ep_2\sum f_I(t){\partial \Psi \over \partial t_I},
$$
where $f_I(t)$ are some functions of the `time' variables $t_I$ and the momentum operator is given by either $p = \ep_1 \del_x$ or $p= \ep_2 \del_x$, depending on the type of brane under consideration. In the refined topological B-model dual to this matrix model, the $t_I$ are identified with certain moduli of the corresponding Calabi-Yau geometry. This makes concrete and elucidates the proposal in \cite{nesh} regarding the existence of the ${\bf  t}$-deformation of the quantum integrable system. In the NS limit we have $\ep_1\ep_2\rightarrow 0$ and the time dependence vanishes, and we simply obtain the time-independent Schr\"odinger equation \req{Hpsisec1} for the $\ep_2$-type brane.

Note however, even though we have found solutions to the time-independent Schr\"odinger
equation in the NS limit, there is no guarantee that the wave-function has no monodromy. 
Some extra quantization conditions are required in order for the wave-functions to be well-behaved under monodromy. 
On the other hand, it was known from \cite{adkmv} that taking branes around the cycles of a Calabi-Yau shifts the dual periods in units of $g_s$ in the usual topological string.  In the NS limit we have $g_s = (-\ep_1\ep_2)^{{1}/{2}}\rightarrow 0$, and the shifts become equivalent to computing derivatives.  Thus
we obtain the statement that taking the brane around the cycles gives the gradient
of the closed string partition function with respect to the closed string moduli. 
This gives an effective way to compute the closed string partition function of the refined topological string theory in the NS limit.

Furthermore, to have a well-defined wave-function we need the wave-function to be single-valued, which by the above argument requires that the exponential of the gradients of the partition function have to be equal to one, leading to the relation \req{NSintcond}.
Therefore, we understand the observation of NS regarding the relation between gauge theory and quantum integrable system as a consequence of the  consistency of open refined topological string theory.
Topological string theory with branes is automatically the relevant system to
study to understand the integrable models in the NS limit, and the geometric engineering of gauge theory tells
us that the same physical quantities can also be computed using target space physics and in particular
the instanton calculus, hence closing the circle of ideas.

Even though the derivation we found was done in the context of matrix models, we conjecture
that it applies to all cases where refined topological strings are defined.  In particular, we argue that one can recover the refined closed string
partition functions in the NS limit, simply by computing the wave-function solution to the Schr\"odinger equation
and seeing what monodromies it picks up as we take it around cycles.  We verify this general
conjecture in many examples.

The organization of this paper is as follows. In section \ref{Rev} we review some earlier results concerning topological strings and its
connection with quantum integrable structures \cite{adkmv}, on which this work heavily relies. This is followed by recalling the various definitions of refined A- and B-model topological strings in section \ref{Refined Topological Strings}.  In section \ref{MMtimedepS} we show that for B-models dual to matrix models, the brane partition functions in the refined topological string satisfy a time-dependent Schr\"odinger equation.  Taking the NS limit of the refined topological string, the time-dependent Schr\"odinger equation reduces to the usual time-independent Schr\"odinger equation as we will show in section \ref{QsurfacesNS}.  We then use this result to explain the observation of NS about the relation between gauge theory partition functions and the Bethe-ansatz for integrable systems.  Sections \ref{MMexamples} and \ref{toric} are devoted to explicit examples illustrating and testing our results and conjectures for the NS limit in the case of genus 0 and genus 1 surfaces. 
In section \ref{lio} we consider B-model branes in Penner type geometries. In this case our time-dependent
Schr\"odinger equation is identical to the BPZ equation for the corresponding degenerate fields of Liouville, and we show that our methods provide a way to derive the $n$-point function by studying the conformal blocks with degenerate insertions.   
Finally, in section \ref{conclusion} we present our conclusions.

Recently there has been an increased interest in the NS limit and several papers on related topics have 
appeared during the course of this work \cite{Marshakov2010,T10,MT10,Dorey:2011pa}.

\section{Review of Quantum Geometry in Topological Strings}
\label{Rev}

In this section we review some elementary features of topological strings on local Calabi-Yau threefolds of relevance to this work. Since topological strings on such backgrounds have been extensively studied, we will be brief.

We will be mainly interested in the B-model topological string and we focus our attention on 
local Calabi-Yau threefolds $X$ which are given by a hypersurface of the form
\be\label{hypersurface}
uv+H(x,p)=0\,,
\ee
where $u,v \in {\C}$ and $x,p \in \C$ or $\C^*$. The classical (\ie, tree level) amplitude of the B-model is encoded via the special geometry of $X$ in the periods of the holomorphic 3-form
$$\omega ={du\over u}\wedge dp \wedge dx\,.$$
For this class of backgrounds, the periods of $\omega$ reduce to residue integrals on $u=0$.  In detail,
if we view the 3-cycles as a fibration of the circle obtained by the rotation $(u,v)\rightarrow (u e^{i\theta}, v e^{-i\theta})$ over a disc $D$, where $u=0$ on the boundary of the disc, \ie, $\partial D\subset \{ H(x,p)=0\}$, then the period integrals reduce to
$$\int_D dp\, dx =\int_{\partial D}p \, dx\,,$$
where the boundary $\partial D$ is a curve on $H(x,p)=0$.  Thus the B-model reduces to the study of the periods of the 1-form
\beq\eqlabel{c1form}
\lambda =p\,dx\,,
\eq
along the 1-cycles of the Riemann surface 
\beq\eqlabel{RSdef}
\Sigma:\, H(x,p)=0\,.
\eq
Hence, we end up with a local version of special geometry which involves a 1-form on a Riemann surface. The pair $(\Sigma,\lambda)$ is often refered to as the spectral curve.

\subsection{Quantum Mechanics and Loop Corrections}

This is the story at tree level.  The natural question is what the quantum corrections do to this picture, \ie, what is the analog of `quantum special geometry' ?  It turns out that an important role is played by the notion of branes.

Consider a B-brane given by the $v$-plane at $u=0$, concentrated on a point $x$ of $\Sigma$, where $p=p(x)$
is fixed in terms of $x$ by the condition $H(x,p)=0$. The {\it classical} partition function $\Psi(x)$ of this brane,
which is a function of its moduli $x$,
was studied in \cite{av00} and was found to be %
\be\eqlabel{bc}
\Psi_{\text{\it class.}}(x)=\exp \left(\frac{1}{ g_s}\int^x p(y)\, dy\right)\,, 
\ee
where $g_s$ is the coupling constant of the topological string.  This structure is very reminiscent of
the WKB approximation to the ground state wave-function, if we identify $H(x,p)$ as the Hamiltonian of the quantum system.  In fact, it
was argued in \cite{adkmv} that on the B-brane phase space the variables $p$ and $x$ do not commute and we have the relation 
%
\be\label{qm}
[p,x]= g_s\,,
\ee
exactly as one has in the usual set-up of quantum mechanics.  Moreover this strongly suggests that the quantum
corrections to the partition function should make $\Psi(x)$ an exact wave-function for the quantum Hamiltonian $H$, \ie, we expect a relation of the form
\beq\eqlabel{Hpsi}
H(x,p)\,\Psi (x) =0\,,
\eq
as the operator realization of the geometric condition  $H(x,p)=0$.  This was proposed in \cite{adkmv}
and checked in various examples.  It was found that for simple cases this is exactly right and gives
the full answer for the quantum corrections to the wave-function. For example, in the case of the (deformed) conifold
$$
H(x,p) =- p^2 + x^2 - \mu\;,
$$
the full partition function of the brane is indeed given by the corresponding energy eigenstate of the harmonic oscillator. 

A motivation for the above commutation relation is the following. 
The brane wrapping the $v$-plane has fields $x,p$ living on it that capture the normal deformations of the brane
as a function of $v$.  The kinetic term for this brane is given by
$$S_{\text{\it B-brane}}={1\over g_s}\int_{\text{\it v-plane}} x{\overline \partial}p+...\,,$$
which leads to the fact that $p$ and $x$ are conjugate variables.

A perhaps more familiar setup to motivate \req{qm} is via mirror symmetry. When $X$ is a mirror to a toric Calabi-Yau manifold $X^*$, the B-brane on the $v$-plane gets mapped to an A-brane on $X^*$ wrapping a non-compact Lagrangian three manifold, $L = \R^2 \times S^1$. The theory on a Lagrangian B-brane is a $U(1)$ Chern-Simons theory, with the classical action 
\be\label{cs}
S_{CS} = \frac{1}{ g_s} \int_L A \, d A\,.
\ee
In the absence of world-sheet instantons, the action \req{cs} is exact. The Chern-Simons path integral on $L$ computes the A-model partition function with the brane inserted \cite{W92CS}. 
Since $L$ is a manifold with a $\T^2$ boundary, the Chern-Simons path integral on $L$ determines a state in the Hilbert space of the theory on $\T^2 \times \R_t$ in a familiar way. As is evident from \req{cs}, when quantizing Chern-Simons theory on $\T^2 \times \R$ with $\R$ viewed as time, the holonomies of the gauge field around the $(1,0)$ and $(0,1)$ cycles of $\T^2$ become canonically conjugate to each other. 
In fact, mirror symmetry directly relates the B-model variables $x$ and $p$ to the complexified holonomies
$$\oint_{S^1_{(1,0)}} A= x\,,\,\,\,\,\,   \oint_{S^1_{(0,1)}} A= p\,.$$   
In general, the world-sheet instantons will correct the classical action, but not in a way that affects this observation. (Namely, they shift $S_{\text{CS}}$ to $S_{\text{CS}} + S_{\text{inst}}$, but the instanton generated terms are independent of time.)   

An example of this is given by the mirror of the A-model on the (trivial) background $X^* = {\C}^2 \times \C^*$, with the $S^1$ in $L$ mapped to the circle in $\C^*$. In this case, the mirror is 
$$H(x,p)=e^p-1\,,$$
where the holonomy around the $S^1$ maps to $x$ by mirror symmetry. In this case, there are no world-sheet instantons, and \req{Hpsi} leads to the trivial result
$$
\Psi(x) =1,
$$
which is the correct answer for the Chern-Simons partition function on $L$. 

A  less trivial example corresponds to the A-model on $X^* = {\C}^3$.
This maps to the local threefold geometry where the spectral curve is given by \cite{hiv00} \footnote{Note that the fact that only terms $\sim e^{n x+m p}$ with $n,m$ being integers appear in the equations for B-model mirrors is related to the fact that $x$ and $p$ are variables mirror to holonomies on $\T^2$, which are clearly periodic.}
$$H(x,p)=e^p+e^x-1=0\;.$$
The equation \req{Hpsi} now leads to the wave-function given by the quantum dilogarithm
$$
\Psi(x) = \prod_{n=0}^\inf \left(1 - e^x q^n\right)\,,
$$
where we defined $q:=e^{g_s}$, as the solution capturing higher loop corrections for the open topological
string. This is mirror to the A-model partition function, when one includes instanton corrections along with the Chern-Simons functional \cite{OV, av00}.

It was found in \cite{adkmv} that this simple picture is also true for all the B-model geometries where $\Sigma$ is a genus zero Riemann surface. In terms of the mirror A-model, this corresponds to geometries with no closed 4-cycles. However, it was also found in \cite{adkmv} that the picture is more complicated if $\Sigma$ is a higher genus Riemann surface. Namely, in general, $\Psi(x)$ is a zero eigenstate of an Hamiltonian $H'(x,p)$ which is not the classical Hamiltonian
$H(x,p)$ but receives $g_s$ corrections.  In other words we have a quantum Hamiltonian ground state
$$H'(x,p)\,\Psi(x) =0, \ \ {\rm where}\ \ H'(x,p) = H(x,p)+\Ocal(g_s)\,.$$
We do expect such quantum corrections, since in general there are already normal ordering ambiguities in replacing the classical variables $p,x$ with the quantum operators.
This shifts the question of the utility of this approach to a better understanding of how to compute corrections to $H$, which was not addressed in \cite{adkmv}, but will  be answered later in this paper.

\subsection{Branes and Period Shifts}
\label{sec2shifts}
Besides the calculation of quantum corrections, there is a further issue to address.  
So far we have only discussed the partition function of the topological
string in the presence of a brane.  How can we obtain the partition function of the closed topological
string in the absence of the brane?

These two issues are in fact closely related. It turns out that the insertion of branes affects the closed string moduli \cite{adkmv}. The meromorphic 1-form \req{c1form} acquires a first order pole with residue $g_s$ at the position $x_0$  where a brane was inserted, \ie, 
$$
\lambda \sim {g_s \over x -x _0} \,dx+ \ldots\,.
$$
This implies that, if we measure
the change in the period of $\lambda$ around the position of the brane, we find that we pick up a period
\be\label{dg}
\oint_{x_0} \!\! \lambda = g_s\,.
\ee
The content of this statement is that topological branes source the topological gravity fields, \ie, they change the background geometry.
This interpretation is in particular supported, in the mirror setup, by the large $N$ duality of the topological A-model on the (singular) conifold
with the resolved conifold $T^*S^3$ \cite{GV99}. Namely, putting $N$  branes on $S^3$ changes the K\"ahler class surrounding
the $S^3$ by $N g_s$.    

In \cite{adkmv} this statement was used to provide a quantum definition of D-branes in the closed string field theory of the B-model. Namely, fluctuations of the one-form $\lambda$ corresponding to quantum fluctuations of complex structures on $X$ are captured by the so-called Kodaira-Spencer scalar field $ \phi$ on $\Sigma$ as
$$
\delta \lambda =  \del {\phi}\,.
$$
Since $\phi$ is a closed string field, its two point functions are proportional to $g_s^2$, \ie, 
$$
\phi(x) \phi(x') \sim  g_s^2 \ln(x-x') + \ldots\,.
$$
Moreover, deformation of the geometry in \req{dg} is consistent with the brane insertion operator $\psi(x)$ being a fermion
$$
\psi(x) = e^{ \phi(x) /g_s},
$$
related to $\phi$ by bosonization. This is also consistent with the period
$$ \oint_{x_0}  \del \phi \;\psi(x_0) =  g_s \psi(x_0) \,.
$$

Furthermore, if we define $\phi(x)$ so that it includes  the classical piece corresponding to $\int^x \lambda = \int^x p\,dx$, then naturally \req{bc} is the classical piece of the insertion of the operator $\psi(x)$ at the point $x$ on the Riemann surface. The fact that the B-brane is a fermion of the Kodaira-Spencer scalar field $\phi(x)$ has been used in \cite{adkmv} to compute the exact B-model amplitudes on some local Calabi-Yau geometries.

The change of moduli due to the insertion of branes means that if we consider a brane/anti-brane pair, and take
one of the branes around a 1-cycle $\gamma$ on $\Sigma$ before bringing it back to annihilate with the other brane, we have changed in the process
the periods of $\lambda$ along any other 1-cycle $\alpha$ by the amount
$$\delta_\g \!\int_\alpha \!\!\lambda =g_s \langle\alpha, \gamma\rangle\,,$$
where $\langle \alpha,\gamma\rangle$ denotes the intersection product of the two 1-cycles.

In particular, if we denote by $a_I$'s the periods of the A-cycles in the Riemann surface $\Sigma$ and decompose the cycle $\gamma$ into a combination of symplectically paired B-cycles, \ie,
$$
\gamma =  \sum m_I B^I\,,\qquad \langle A_I, B^J \rangle = \delta_{I}^J\,,
$$
we find that the closed string partition function is shifted by
$$Z_{closed}(\vec a)\ \rightarrow\  Z_{closed}(\vec a+ g_s \vec m)\,,$$
where $\vec a$ denotes the collection of the K\"ahler moduli. 
This implies that once we know the partition function of topological
strings in the presence of branes, we can in principle find the closed string partition function by its variation with respect to the closed string moduli, 
at least for shifts of $a_I$ by integral multiples
of $g_s$.

Interesting as these statements are, they leave a gap in their applicability.
This is because for general cases, we cannot compute the brane wave-function $\Psi(x)$ since we do not know how to compute the quantum corrections to the quantum Hamiltonian $H(x,p)$.  It turns out that these problems get remedied when we consider a one-parameter deformation
of topological strings, inspired by the related deformation of gauge theory partition functions \cite{nekrasov}.

\section{Refined Topological Strings}
\label{Refined Topological Strings}

There is an interesting one-parameter extension of topological strings in the case of non-compact Calabi-Yau backgrounds. Denoting this parameter by $\beta$, the extension can be combined with the usual string coupling constant $g_s$ into two parameters $\epsilon_1,\epsilon_2$ defined by
\beq\eqlabel{epbetags}
\epsilon_1=-g_s\sqrt{\beta},\qquad \epsilon_2=g_s/\sqrt{\beta} \; .
\eq
The standard topological string corresponds to the special case $\beta=1$, where we have 
\beq\eqlabel{topstringeps}
-\epsilon_1=\epsilon_2 = g_s\,.
\eq
However, unlike the usual topological string theory, we do not yet have a world-sheet description of the refined topological string with generic extra parameter $\b$. Nevertheless, in various cases we do have a definition in terms of space-time physics. Let us now review the different definitions.

In the topological A-model we can give a general definition for this one-parameter deformation in terms of the partition function of target space physics. Assuming mirror symmetry is general, this would be a working definition for the topological B-model as well. The extension in the A-model case is accomplished by lifting to M-theory. The addition of an extra dimension closely parallels the construction of Nekrasov of $\Omega$-backgrounds in ${\cal N} = 2$ gauge theories. In particular, when the Calabi-Yau geometry engineers a four-dimensional gauge theory, the computation of the target space partition function can be captured by the instanton calculus of \cite{nekrasov}. 
More generally, the one-parameter extension of the topological string has a space-time analogue which is a global ${\cal N} = 2 $ supersymmetric system in four dimensions. 

In the next subsection we will first review the definition of A-model topological strings in terms of M-theory and show how this definition can be extended to the cases where the Calabi-Yau is non-compact. Then we will relate this to the spectrum of BPS states in five dimensions. We then connect this picture to Nekrasov's definition in the special case where the Calabi-Yau engineers a five-dimensional gauge theory and briefly comment on this extension for a general global ${\cal N}=2$ system. Finally, we discuss a direct B-model definition of this extension when the topological string can be realized in terms of matrix models.

\subsection{The A-model, its M-Theory Lift and Deformation}
\label{refMtheory}
The partition function of the topological A-model on a Calabi-Yau $X$ is equal to the partition function of M-theory on the space \cite{Dijkgraaf:2006um}
$$
X \times S^1 \times TN\,,
$$
where $TN$ is the Taub-NUT space, viewed in complex coordinates $z_1,z_2$ .  Furthermore, the Taub-NUT space is twisted along the $S^1$, in the sense that going around the circle rotates the coordinates $z_1,z_2$ by 
$$
z_1 \to e^{i\ep_1} z_1,\qquad z_2 \to e^{i\ep_2} z_2\;.
$$
In order to preserve supersymmetry we need 
$\ep_1 + \ep_2 = 0\,,$
and we find \req{topstringeps}. The refined topological string is obtained by relaxing this constraint on the $\epsilon_\alpha$. However, to do so while preserving supersymmetry, we need an extra $U(1)_R$ symmetry acting on $X$. This can only be accomplished with $X$ being non-compact, a condition we will assume from now on. 

This partition function is not easy to compute, but can be related to the spectrum of BPS particles in five dimensions \cite{GV982,hiv}.
Namely, the corresponding M-theory partition function  simply computes the BPS indices
$$
N_{j_L, j_R}^{{\vec d}}
$$ 
of BPS states of M2 branes wrapping 2-cycles of $X$ of class ${\vec d} \in H_2(X, {\mathbb Z})$, and with
$SO(4) = SU(2)_L \times SU(2)_R$ spin quantum numbers $j_L, j_R$. 
Decomposing the 
corresponding field into modes
$$
\Phi(z_1,z_2) = \sum_{n_1,n_2} a_{n_1,n_2} z_1^{n_1} z_2^{n_2},
$$
the $U(1)_L \times U(1)_R \subset SU(2)_L\times SU(2)_R$ acts on this by
$$
(z_1,z_2) \rightarrow (e^{i (\theta_L+  \theta_R)} z_1,  e^{i (-\theta_L+ \theta_R)} z_2).
$$
The BPS partition function is defined as a trace in the Hilbert space ${\cal H}_{BPS}$ of a gas of spinning M2 branes, weighted by their total spin and charges:
$$Z_{BPS} = {\rm Tr}_{{\cal H}_{BPS}} (-1)^{2(m_L+m_R)}q_L^{2 m_L} q_R^{2 m_R}\, e^{-\vec a\cdot \vec d}\,\;,
$$
where we have used $(m_L,m_R)$ to denote the spin content $(j_L^3,  j_R^3)$ of the highest spin state in a given BPS multiplet. 
In this case, the Hilbert space is just the Fock space of a single M2 brane and hence the partition function takes the form\footnote{Strictly speaking, this statement is true only with large three-form fluxes turned on \cite{Denef:2007vg}. See also \cite{Aganagic:2009kf}. }
$$
Z_{BPS}= \prod_{{\vec d}, \,j_L,\, j_R}\prod_{m_L = -j_L}^{j_L}\prod_{m_R = -j_R}^{j_R} \prod_{n_{1}, n_2 \geq 0}
\left(1- e^{-\epsilon_1 (m_L+m_R+\frac{1}{2} + n_1)}e^{\epsilon_2 (m_L-m_R+\frac{1}{2} + n_2)} e^{-\vec a \cdot \vec d}\right)^{ N_{j_L, j_R}^{{\vec d}}}\;,
$$ 
where $q_L = e^{-\frac{\ep_1-\ep_2}{2}}$, $q_R = e^{-\frac{\ep_1+\ep_2}{2}}$ and  $\vec a=(a_1,a_2,\dots)$ the set of K\"ahler parameters. Note that in addition to the intrinsic angular momenta of the particle in five dimensions, the orbital angular momenta   contribute to $j_{L,R}$ as well. 

In the special case of $q_R=1$ we effectively trace over the $SU(2)_R$ quantum numbers. In this limit, the BPS partition function equals the topological string partition function, and we have the usual relation between the topological string and M-theory amplitude.  For the refined version of the topological string we define
$$
Z_{\text{\it top}}(\vec a;\ep_1,\ep_2) =Z_{BPS}(\vec a;\ep_1,\ep_2)\,.
$$ 

Equivalently, we could have obtained $ \Fcal_{BPS} = \log Z_{BPS}$ from a one-loop, Schwinger type computation 
which involves integrating out M2 brane particles in a gravi-photon background parameterized by $\epsilon_1$ and $\epsilon_2$ that is no longer self-dual \cite{hiv}. 
This gives ${\Fcal}_{BPS}$ as 
\beq\eqlabel{GVref}
\begin{split}
\Fcal_{BPS}(\vec a;\epsilon_1,\epsilon_2)=&\sum_{\vec d}\sum_{j_L,j_R=0}^\infty\sum_{k=1}^\infty N_{j_L,j_R}^{\vec d}\, \frac{e^{-k \vec a \cdot {\vec d}}}{k}\\ 
&\times \frac{\Bigl(e^{-(\e_1+\e_2) k j_R}+\ldots+e^{(\e_1+\e_2) k j_R}\Bigr)\Bigl(e^{-(-\e_1+\e_2) k j_L}+\ldots+e^{(-\e_1+\e_2) k j_L}\Bigr)}{(e^{k\frac{\e_1}{2}}-e^{-k\frac{\e_1}{2}})(e^{k\frac{\e_2}{2}}-e^{-k\frac{\e_2}{2}})}\,.
\end{split}
\eq
For some, but not all choices of $X$, this will also have a gauge theory interpretation in terms of a five-dimensional gauge theory on a circle.

\paragraph{Branes in the Refined A-model Topological String}
\noindent One can extend the above formulation to include A-branes. The usual A-type brane lifts in M-theory to a M5 brane wrapping the mirror Lagrangian 3-cycle $L$ in $X$, and the ${\C} \times S^1$ subspace of $TN\times S^1$  \cite{OV}.   A similar consideration also applies to the refined version of topological strings.  
In this case we have {\it two} inequivalent types of branes, depending on which of the cigar subspaces of $TN$, {\it i.e.},  the $z_1$- or the $z_2$-plane, the $M5$ brane wraps.
For general $\epsilon_\alpha$'s, the symmetry between $z_1$ and $z_2$ is broken and the two types of branes are no longer on equal footing.
At low energies, the theory on the brane has ${\cal N}=2$ supersymmetry in three dimensions. After introducing A-branes, in addition to M2 brane particles wrapping closed holomorphic 2-cycles in $X$, there are also M2 branes wrapping holomorphic disks in $X$ and ending on $L$. The branes break the local $SO(4)$ rotation symmetry to $SO(2)_1\times SO(2)_2$. Combining the $SO(2)_L$ subgroup of it together with the $SO(2)$ R-symmetry, we still have an $SO(2)_L\times SO(2)_R = U(1)_L \times U(1)_R$ symmetry available \cite{gsv}.
In the presence of the branes we consider the BPS partition function of open and closed M2 branes, keeping track of their spin and the relative homology class in $H_2(X, L)$.  The latter corresponds to the fact that the open M2 branes are charged under the world-volume (magnetic) gauge field on the M5 brane (see \cite{OV} for more details). In this way we can define the open BPS partition function, analogously to the closed one.
%
We refer to the partition function in the presence of the two different types of branes, corresponding to the
M5 brane wrapping the $z_{\alpha=1,2}$ plane, as $\Psi_{\alpha}$.

\subsection{$\Omega$-Deformation of ${\cal N}=2,$ $d=4$ Theories}

The so-called $\Omega$-background for four-dimensional ${\cal N} =2$ supersymmetric theory, where the parameters $\epsilon_{1,2}$ correspond to the equivariant $U(1) \times U(1)$ action on the Euclidean space-time ${\C}^2 \cong {\R}^4$, has been considered in \cite{nekrasov, no, nw10}, based on earlier work \cite{Moore:1997dj,lns2}. More precisely, one considers a deformation of the theory where one replaces the adjoint-valued scalars $\Phi$ by
\be\label{eq}\Phi \rightarrow \Phi + \sum_\alpha \epsilon_\alpha z_\alpha {D\over D z_\alpha}\,,
\ee
corresponding to adding generators of infinitesimal rotation along the two complex planes of ${\C}^2$,  with coordinates $z_{1,2}$, accompanied by an R-symmetry twist. Since $\Phi$ gives masses to charged fields in the theory, this deformation effectively adds a mass to the modes of the fields in the theory according to their transformation properties under the $U(1)\times U(1)$ action. The partition function of the theory in this background defines a refined partition function $Z_{{\cal N}=2}(\epsilon_1,\epsilon_2)$.  

When the ${\cal N}=2$ theory in question is a gauge theory, the $\Omega$-background allows one to compute the partition function explicitly by  performing integrals over the instanton moduli spaces. Moreover, if we consider a five-dimensional gauge theory 
with an additional $S^1$, the partition function can also be computed using the instanton calculus.  
Furthermore, this maps to our M-theory construction above in the cases where the local Calabi-Yau $X$ engineers the corresponding 5d gauge 
system, and thereby makes contact with the refined A-model topological string. 
In fact it was explicitly verified in \cite{iqka,Iqbal:2003ix} that in the special case \req{topstringeps}, the partition function $Z_{{\cal N}=2}(\e_1,\e_2)\rvert_{-\ep_1=\ep_2=g_s}$ agrees with the topological string partition function $Z_{\text{\it top}}(g_s)$ on the corresponding Calabi-Yau manifold. Furthermore it was checked that a refined version
of the topological vertex not only reproduces the $SU(2)_L\times SU(2)_R$ quantum numbers of BPS states, but
it also agrees with the full Nekrasov partition function of the ${\cal N}=2$ theory \cite{ikv}.

This viewpoint on deformation also extends to the open topological string. The A-branes can be given a purely gauge-theoretic formulation as surface operators, studied recently  in \cite{Gaiotto:2009fs, Alday:2009fs,Dimofte:2010tz,Kozcaz:2010af}, for example. For general deformation $\ep_\alpha$'s, the symmetry between $z_1$ and $z_2$ is broken and we obtain two types of branes. The breakdown of symmetry can already be seen at the classical level.

From the string theory perspective, the surface operators are described by the branes of the theory.
The world-volume theory on the brane wrapping the $z_{1,2}$-plane starts out as an ${\cal N}= (2,2)$ supersymmetric theory in $d=2$ with a superpotential $W(x)$, where $x$ is a chiral superfield.
Since the equivariant action is in essence a kind of Kaluza-Klein reduction that effectively gives the $z_\alpha$ plane a volume proportional to $1/\epsilon_\alpha$, the classical partition function of the brane wrapping the $z_\alpha$ plane becomes
$$
\psi_{\alpha,\text{\it class.}}(x)=\exp \left(\frac{W(x)}{\epsilon_\alpha}\right)\quad,\quad \alpha=1,2\;.
$$
In the case of the B-model topological string, as discussed in section \ref{Rev}, the corresponding superpotential
is identified with
$$
W(x) = -\int^x p(x')dx'\,,
$$
leading to a WKB-type wave-function
\beq\label{bce}
\psi_{\alpha}(x)=\exp \left(-\frac{1}{\epsilon_\alpha}\int^x p(x')dx'\right)\,. 
\ee
for the two types of branes wrapping the $z_\alpha$-plane.

\subsection{The $\b$-Ensemble as a Refinement of the Topological B-model}
\label{MM1}

The above discussion was geared towards A-model topological strings.  
By employing mirror symmetry it also gives us an answer for the B-model in principle. 
However, given the simplicity of the B-model, it is convenient to give a direct definition of the refined B-model.  
Indeed a proposal for such a definition has been put forth in \cite{DV09} in terms of a certain deformation of matrix models. 

Large $N$ matrix models provide an alternative description of topological B-models on a class of geometry
and an excellent testing ground for the ideas we just reviewed. In view of its importance in the rest of the discussion, we will now review the relation between random matrices and topological strings. We will start by first reviewing the usual B-model and focus on the refined case afterwards.

For the type of geometries described by the local curve
\beq\label{st}
H(x,p) = -p^2+ W'(x)^2\;
\ee
with $W(x)$ a degree $g+2$ polynomial, there is a conifold singularity near each critical point of the potential $W(x)$ which we can blow up into a $\P^1$.  The B-model topological string in this background has a matrix model description \cite{DV02}. Its partition function is given by
\beq\label{mn}
Z = \int d\Phi \, e^{-\frac{2}{g_s}  {\rm Tr} W(\Phi)} \;,
\ee
expanded around a given distribution $N_I$ of eigenvalues of the matrix $\Phi$ among the $g+1$ critical points of the potential $W(x)$. 

This matrix model can be derived from the B-model \cite{DV02, adkmv} as follows. If we cut the geometry into two halves, corresponding to writing \req{st} as
$$
-H(x,p) = \big(p+ W'(x)\big) \big(p-  W'(x)\big)=H_L(x,p) H_R(x,p),
$$
the branes on the ${\P^1}$ are obtained by gluing non-compact branes on $H_{L,R}(x,p)=0$ across their boundary. 
From $H_L$, with branes inserted at $x=z_i$, for $i=1,\ldots N$ we get
\beq\label{corr}
\left\langle \psi(z_1) \psi(z_2) \ldots \psi(z_N) \right\rangle_{ L} = e^{- {1\over g_s} \sum_i W(z_i)} \;\prod_{j<k} (z_j-z_k) \,.
\ee
The interaction term comes from the two-point functions of $\psi(z_i)$ with each other, while the potential term is the wave-function of a single D-brane as we explained before.  
%
Since this is a genus zero Riemann surface, the result is exact as we will argue later.
Similarly, from $H_R(x,p)$ we end up getting another copy of this, as the orientation of the branes is naturally opposite. Setting the values of $z_i$ equal on both sides and integrating them, we indeed obtain the partition function \req{mn}, which reads in the eigenvalue basis
$$
Z = \int d^N z \prod_{i<j} (z_i - z_j)^2  e^{- {2\over g_s} \sum_i W(z_i)}\,.
$$

It was conjectured in \cite{civ01,DV02} that the geometry with the branes go through a transition to a smooth geometry
where the $\P^1$'s are replaced by $S^3$'s with size
$$
\mu_I = g_s \,  N_I\;. 
$$
More precisely, we now have the geometry given by  
\beq\label{after}
H(x,p) = -p^2+ W'(x)^2+f(x)\;,
\ee
where the correction (deformation) $f(x)$ term is a degree $g$ polynomial and determined by the distribution 
$$
N = N_1 + \ldots + N_{g+1}\,,
$$
of the $N$ branes over the $g+1$ ${\P}^1$'s given by critical points of $W(x)$. 
The open-closed duality states that the matrix model, in the large $N$ 't Hooft limit 
$$
N \to \infty, \; g_s \to 0\;,\; N g_s \text{  fixed}\,,
$$
gives a description of the B-model on  the geometry \req{after} after the transition.
The curve 
\be\label{HDV}
\Sigma: H(x,p)=-p^2+ W'(x)^2+f(x)= 0 
\ee
is now the spectral curve of the matrix model, from which the higher $g_s$ correction of the matrix model partition function can be computed \cite{EO07}. 

From the same argument, we can also deduce that the insertion of a non-compact brane corresponds to the insertion of the characteristic determinant on the matrix model side and leads to the brane partition function 
\beq\label{dt}
Z_{\text{\it brane}}(x) = \langle \psi(x) \rangle= e^{-\frac{1}{g_s} W(x)}\int d\Phi \, e^{-\frac{2}{g_s}  {\rm Tr} W(\Phi)} \det(\Phi -x)\,. 
\ee

%
%

The generalization of the above picture to the refined background is proposed in \cite{DV09}. 
For $\beta\neq 1$, or $\e_1+\e_2\neq 0$, the matrix model obtains a different measure while retaining the same potential. 
The generalization simple changes the power of the Vandermonde determinant to $2\beta$,
together with a rescaling of the coupling constant in front of the potential $W$, \ie, 
$$
Z = \int d^N \!z \prod_{i < j} (z_i - z_j)^{2\beta}e^{-2  \frac{\sqrt\beta}{ g_s} \sum_i W(z_i)}\,.
$$
In terms of the variables $\epsilon_1,\epsilon_2$ this translates into
\beq\label{rmm}
Z = \int d^N \! z \, \prod_{i < j} (z_i - z_j)^{-2{\epsilon_1}/{\epsilon_2}} \,e^ {-\frac{2}{\epsilon_2} \sum_i W(z_i)}\,.
\ee
In particular, for logarithmic potentials relevant for the discussion on relations to ${\cal N}=2$, $d=4$ gauge theories, the $\b$-ensemble matrix model integral takes the form of a Coulomb gas representation of the conformal blocks of Liouville theory with background charge given by $\beta$. 

This deformation can also be understood from our space-time perspective. Namely, if we consider branes wrapping the $z_\alpha$ plane, in the space time ${\C}^2$,  rotated by $\epsilon_\alpha$, the classical partition function becomes \req{bce}, instead of \req{bc}. So, in deriving the matrix model, as we reviewed in section \ref{Rev}, we expect to simply replace the fermion $\psi(x) = \exp( \phi(x)/g_s)$,  by the operators
\beq\label{newbrane}
\psi_{\alpha}(x) = \exp( \phi(x)/\epsilon_\alpha),\qquad \alpha=1,2.
\ee
On the other hand, the scalar field $\phi(x)$ still has the same correlation functions as before (this is the essence of the Coulomb gas formalism), which is  consistent with the fact that the quantum mechanics of the closed string is unchanged, as we will later discuss in the context of holomorphic anomaly. This means that the correlator \req{corr} computing the partition function of branes on the halved geometry changes to
\beq\label{corr}
\left\langle \psi_{2} (z_1) \psi_{2} (z_2) \ldots \psi_{2} (z_N) \right\rangle_L = \prod_{i<j} (z_i-z_j)^{g_s^2/\ep_2^2}  e^{- {1\over \ep_2} \sum_i W(z_i)}\;,
\ee
which leads to the partition function \req{rmm} upon gluing, using the fact that $g_s^2 = - \ep_1\ep_2$.


From the saddle point equation
$$
W'(z_i) = -\e_1 \sum_{j\neq i}\frac{1}{z_j-z_i}\;,
$$
we see that the relevant 't Hooft coupling is given by
$$
\mu= \e_1 N \;.
$$
We see that the amount by which the branes change the geometry around them now depends on the type of the brane. More generally, from \req{newbrane}
we expect that inserting an $\ep_{1}(\ep_2)$ brane at a point $x_0$ on the Riemann surface deforms the geometry by $\ep_2(\ep_1)$.
\be\label{ba}
\oint_{x_0} \del \phi \; \psi_{\alpha}(x_0) = \frac{g_s^2}{\epsilon_\alpha} \psi_{\alpha}(x_0).
\ee

We can also easily describe non-compact D-branes for this matrix model. The two different kinds of branes 
$$
\psi_{\alpha}^*(x) = \exp(-\phi(x)/{\ep_{\alpha}}),
$$
correspond in the matrix model language to the operators 
$$
e^{\frac{1}{\epsilon_\alpha}W(x)}\cdot \text{det}(x-\Phi)^{\e_1/\epsilon_\alpha} \quad, \quad \alpha =1,2, 
$$
respectively. 
Such an explicit realization of branes in the matrix model allows us to study them very directly.  We will come back to this theory in the next section, where we will derive the exact, time-dependent Schr\"odinger equation that they satisfy.

Before we go on, note that we can rewrite the partition function \req{rmm} as
\beq\eqlabel{ZMMforbaby}
Z= \int d^Nz \, \prod_{i< j} (z_i - z_j)^2 \,e^{-\frac{1}{\e_2} \big(2\sum_i W(z_i) + 2(\e_1+\e_2) \sum_{i< j}\log(z_i-z_j)\big)}
\eq
and view the change of the measure from the usual Vandermonde squared as adding a non-local operator with a coupling constant 
$$\hbar:=\e_1+\e_2\;.$$ 
Writing the $\beta$-ensemble partition function as in \req{ZMMforbaby}, makes manifest that apart from the usual genus expansion in $g_s$, the free energy ${\cal F}=\log Z$ has another expansion in terms of the parameter $\hbar$ defined above. We end up with the double expansion \footnote{Note that the refined matrix model free energy \req{FMMgsh} generally possesses an expansion into {\it even} and {\it odd} powers of $\hbar$. 
From the M-theory perspective reviewed in section 2, we expect that the refined topological string partition function actually possesses an expansion into {\it even} powers of $\hbar$ only. This simply comes from the fact that the BPS states fit into complete spin multiplets \req{GVref}. Apparently, the $\beta$-ensemble breaks this symmetry. However, as anticipated in \cite{kw2}, the symmetry can be restored via appropriate redefinition of the (deformation) parameters as we will see later both in the general discussion and in the explicit examples we will consider.}
\beq\label{FMMgsh}
{\cal F}(\vec a;\e_1,\e_2)= \sum_{g\geq 0 , \ell \geq 0} {\cal F}^{(g,\ell)}(\vec a)\, g_s^{2g-2} \hbar^{\ell}  \;.
\eq
The expansion (\ref{FMMgsh}) and its recursion relation has been studied in \cite{Chekhov:2005rr,Chekhov:2006rq,Brini:2010fc}.

\subsection{Refinement, Topological Strings and Quantum Mechanics}
\label{sec3shifts}

One striking aspect of topological string theory, is that quantum mechanics makes two independent, though related, appearances in the theory. On the one hand, as we reviewed above, the {\it open} topological string partition function is a wave-function with the Riemann surface as the level set of the Hamiltonian. As we will see later, this continues to be true after we turn on the two independent parameters $\ep_{1,2}$. On the other hand, the holomorphic
anomaly of \cite{BCOV2} implies that
the {\it closed} topological string partition function is also a wave-function. In other words, it is a state in the Hilbert space obtained by quantizing $H^3(X, {\C})$ \cite{W93}. This remains true for the refined topological string \cite{kw1}.
As we will explain, this fact can be understood from  the way branes deform the geometry and the monodromy transformation of the brane partition function. Moreover,  in section \ref{QsurfacesNS} we will see how the monodromy properties of the open topological string partition function can be used to derive differential (or difference) equations that the closed string partition satisfies.

Given a symplectic basis of $H_3(X, {\mathbb Z})$ with
$A_I \cap B^J = \delta_{I}^J\,,
$
the periods of the holomorphic three form  $\omega$ parameterize the phase space $H^3(X, {\C})$. In our case, this is equivalent to periods of the one-form $\lambda$ on the Riemann surface $\Sigma$
$$
a_I = \oint_{A_I} \lambda\,, \qquad a_D^J = \oint_{B^J} \lambda\,.
$$ 
Classically, $a_D^I$ and $a_J$ are not independent, but satisfy the special geometry relation
\beq\label{sc}
a_D^J = \frac{\partial}{\partial a_J} {\Fcal}^{(0)}
\ee
in terms of the genus zero topological string amplitude $\Fcal^{(0)}$. On the exact, unrefined, topological string partition function
$$
Z_{\text{\it top}}(\vec a;g_s)= \exp\left(\sum_g {\cal F}^{(g)}(\vec a)\, g_s^{2g-2}\right)\,,
$$
the periods $a_I, a_D^J$ are realized as canonically conjugate operators,
\be\label{dual_a_commutation}
[{\hat a}_D^J, {\hat a}_I] = g_s^2 \delta_{I}^{J},
\ee
where we used hats to distinguish the operators from their expectation values. In particular, \req{sc} is a semiclassical approximation to this.

The above quantum equation as well as its generalization to the refined topological strings can be derived by considering the way the branes deform the geometry.
As we have seen in section \ref{MM1}, an $\epsilon_\alpha$-brane deforms the period of $\del \phi$ around any cycle
surrounding it by  (\ref{ba}).

Consider creating an $\epsilon_\alpha$-brane/anti-brane pair at a point on $\Sigma$, and taking one of the branes around a cycle $\gamma$ before annihilating with each other. 
Taking the branes around $\gamma_A= \sum_I \ell^I A_I$ does not change the expectation value 
$$a_I = \oint_{A_I} \del \phi\,.$$ 
(Note that $\oint_{\gamma} \del\phi$ is the quantum generalization of $\oint_{\gamma} p dx$.) However, such a monodromy around $\gamma_A$ changes the phase of the partition function by
$$
{\cal M}_{\gamma_A}:\quad  Z_{\text{\it top}}(\vec a)\rightarrow \exp\left(\frac{1}{\epsilon_\alpha} \sum_I \ell^I a_I \right) Z_{\text{\it top}}(\vec a)\,,
$$
since this change of phase is induced by the change of phase of 
$$\langle ... \exp(-\phi(x)/\epsilon_\alpha)\rangle\,,
$$
as we transport the brane around.
This is to be contrasted with the situation if we take the branes around a $\gamma_B=\sum_I m_I B^I$ cycle
$$
{\cal M}_{\gamma_B}:\qquad Z_{\text{\it top}}(\vec a) \rightarrow Z_{\text{\it top}}(\vec a+ \frac{g_s^2}{\epsilon_\alpha} \vec m) 
=  \exp\left(\frac{g_s^2}{\epsilon_\alpha} \sum_I {m_I }{\del \over \del a_I}\right)  Z_{\text{\it top}}(\vec a)\,,
$$
as a consequence of \req{ba}, generalizing the shift reviewed in section \ref{sec2shifts}. Now, consider writing $Z$ in the terms of the dual variables $a_D^I$ associated with the B-cycles instead. 
In this dual basis the monodromy around the cycle $\gamma_B$ acts on $Z$ as a multiplication operator by  
$\exp( \sum_I {\frac{1}{\epsilon_\alpha}m_I a_D^I})$, and that around the cycle $\gamma_A$ becomes a shift operator. 
The consistency of the two dual pictures requires $a_D^I$ to be realized as   
$$
a_D^I = g_s^2{\del \over \del a_I}
$$
acting on the partition function in the $a_I$-basis. We conclude that the commutation relation \req{dual_a_commutation} between the operators $a_I, a_D^J$ also holds for the refined topological string, and therefore the closed string partition function $Z$ is indeed a wave-function on $H^3(X, {\C})$ for arbitrary  $\epsilon_{1,2}$. The only effect of the $\beta$ deformation is to change the unit of the shift and hence the form of the wave-function. 
This is consistent with the observation made in \cite{kw1,kw2} that in known cases the refined topological string partition function still satisfies the holomorphic anomaly equation of \cite{BCOV2}.

\section{Matrix Models and Schr\"odinger Equations}
\label{MMtimedepS}
In this section, we will show that we can derive from matrix models a 
{\it  multi-time} dependent Schr\"odinger equation for arbitrary $\beta$, satisfied by a brane probing a Riemann surface (\ref{HDV}),
with 
\beq\eqlabel{WDV}
W(x) = \sum_{n=0}^{g+2} t_n x^n
\eq
and $f(x)$ a polynomial of degree $g$. We will sometimes take $g$ to infinity, so that $W(x)$ becomes a formal sum. 
The relation between quantum geometry and the $\beta$-ensemble matrix model has been discussed in \cite{EM08,Chekhov:2009mm,Chekhov:2010zg} in terms of the resolvent of the matrix model.  
Here, for us we the natural object to study is the brane partition function for which the quantum geometry of the matrix
model becomes manifest.

We will also see that the time-dependent Schr\"odinger equation satisfied by the brane wave-function can be rephrased in the form of the BPZ equation
\cite{Belavin:1984vu} satisfied by a correlation function in two-dimensional CFT with a degenerate operator insertion, similar as for surface operators in the Liouville context \cite{Marshakov2010}.

\subsection{Time-Dependent Schr\"odinger Equation}
\label{Time-Dependent Schrodinger Equation}

From section \ref{MM1}, inserting an $\epsilon_\alpha$-brane, the refined topological string partition function on this geometry becomes 
\be\label{bmm}
Z_\alpha(x) =e^{\frac{1}{\ep_\alpha} W(x)}\int d^N \!z \prod_{i < j} (z_i - z_j )^{-2\ep_1/\ep_2}
\prod_i(x-z_i)^{\ep_1/\epsilon_\alpha} \;e^{-  \frac{2}{ \ep_2} \sum_i W(z_i)}\,.
\ee
To derive the equation satisfied by (\ref{bmm}), 
consider differentiating it with respect to $x$. For simplicity of the derivation, we will remove the classical piece $e^{\frac{1}{\epsilon_\alpha}W(x)}$ for the time being and restore it later. 
Moreover, we will denote by $h =\ep_1/\epsilon_\alpha$ the power of the determinant  $\text{det}(x-\Phi)=  \prod_i(x-z_i)$. Different values of $h$ correspond to different types of branes. 

It is easy to see that differentiating is the same as inserting the following function inside the intergral
%
\beq\notag
\begin{split}
\frac{\del^2}{\del x^2} &= \sum_{i,j} \frac{h^2}{(x-z_i)(x-z_j)}  -  \sum_{i} \frac{h}{(x-z_i)^2} \\
&=  \sum_{i< j} \frac{2h^2}{(x-z_i)(z_i-z_j)}  +  \sum_{i} \frac{h^2-h}{(x-z_i)^2}\,.\\
\end{split}
\ee
At the same time, from the loop equation 
\begin{equation}\label{ward1}
0=\sum_{i=1}^N \int  d^Nz\, \frac{\pa}{\pa z_i} \left( \, \frac{1}{x-z_i}{\cal O}(z)\,e^{-\frac{2}{\e_2} \sum_j W(z_j)}
\prod_{j < k}(z_j-z_k)^{-2\ep_1/\ep_2}  \right)\,
\end{equation}
with the choice of operator ${\cal O} =\prod_{i}(x-z_i)^{h}$, we obtain the identity  
$$
\sum_{i}\frac{1-h}{(x-z_i)^2} -\frac{2}{\ep_2} \left( \sum_i \frac{W'(z_i)}{x-z_i} +\ep_1 \sum_{i<j} \frac{1}{(x-z_i)(x-z_j)}\right)=0
$$
under the integral sign.
As we will mention in more details later, the above loop equation has the meaning of the Ward identity of a quantum symmetry of the matrix theory. 
In order to have a simple operator equation, we would like to cancel the terms $\sum_i\frac{1}{(x-z_i)^2} $ and 
$\sum_{i\neq j} \frac{1}{(x-z_i)(z_j-z_i)}$ and the above equations show that this is possible exactly when $h$ takes the two values
$h= \epsilon_1/ \epsilon_\alpha$ corresponding to our two types of branes!  

Reincorporating the classical piece $e^{\frac{1}{\epsilon_\alpha}W(x)}$ and 
putting everything together, we find that $Z_\alpha$ satisfies an operator equation (under the integral) of the form

\be\label{int}
-\epsilon_\alpha^2 \pa_x^2+ W'(x)^2 +\epsilon_\alpha W''(x)  +f(x) =0\,,
\ee
with

$$
f(x) =2 \ep_1 \sum_{i=1}^N \frac{W'(z_i) -W'(x)}{z_i-x}\,.
$$ 
The effect of inserting $f(x)$ is the same as acting by a linear differential operator
$$
\hat f(x) = g_s^2 \sum_{n=0}^{g} x^n \partial_{(n)}\,,
$$
with
$$ \partial_{(n)} =  \sum_{k=n+2}^{g+2} k t_{k} {\del\over \del t_{k-n-2}}\,.$$
Here we differentiate the matrix model partition function with respect to the coefficients $t_k$ of the potential. We have also set  
$
\frac{\pa}{\pa t_0}  = -\frac{N}{2\e_2}.
$

In summary,  we find that the brane partition functions $Z_\alpha(x)$ satisfy a linear differential equation
\be\label{differential}
\left( -\epsilon_\alpha^2 \pa_x^2+ W'(x)^2 +\epsilon_\alpha W''(x)  +\hat f(x)  \right)Z_\alpha(x;t) =0\,,\quad \alpha=1,2\;.
\ee
We emphasize that we have not taken any limits here -- the equation is exact. 

This equation is in fact a multi-time dependent Schr\"odinger equation, with the Riemann surface playing the role of a time-dependent Hamiltonian. In order to see this, we can proceed as follows. To begin with, $Z_\alpha(x)$ contains both open and closed string contributions, since it corresponds to an unnormalized expectation value. Correspondingly, the time-dependent Schr\"odinger equation we got has no reference to filling fractions, or the choice of the background, as they do not enter the matrix integral explicitly. For many purposes, the normalized expectation value $\Psi_\alpha(x)$ is a more natural quantity
$$
{\Psi_\alpha(x)} = \frac{Z_\alpha(x)}{Z}\,.
$$
Here $Z$ is the matrix model partition function without the brane, in the sector corresponding to specific filling fractions
$$
N \rightarrow (N_1, \ldots, N_{g+1})\,,
$$ 
around which we expand in the 't Hooft expansion, $Z = Z(\mu_I; \ep_{1},\ep_2)$ where $\mu_I =  N_I\ep_1$ are the 't Hooft couplings. 
The normalization induces explicit dependence on the background.
It follows that ${\Psi}_\alpha(x)$ is the purely open string partition function and satisfies
\beq\label{tdse}
\Bigl(-\epsilon_\alpha^2  \frac{\del^2}{\del x^2}  + W'(x)^2 
+ f(x)+ g_s^2 \sum_{n=0}^{g}  x^n \partial_{(n)} \Bigr){ \Psi}_\alpha(x) = 0\,,
\ee
where
\beq\label{f}
f(x) = \sum_{n=0}^{g} x^n b_n\,
\ee
is now a polynomial with some coefficients $b_n$ that parameterize the complex structure moduli, and thus implicitly contain the choice of the background. More precisely, writing $\log Z = {\cal F}_{closed}/ g_s^2$, we have that
\beq\label{bb} b_n  =  \partial_{(n)} {\cal F}_{closed} + \ep_\alpha(n+1)(n+2) t_n. 
\ee
Notice that we have incorporated the term proportional to $W''(x)$ into our definition of the polynomial $f(x)$. This shift has origin in the ordering ambiguity in quantization and corresponds to the $\tfrac{1}{2}\hbar$ shift in the ground state energy of the harmonic oscillator when we consider the Gaussian potential. 

To understand the meaning of this consider the planar limit
$$\ep_1 \rightarrow 0\quad,\quad N_I\rightarrow \infty \quad\text{with} \quad \mu_I = N_I\ep_1\quad\text{fixed}\;.$$  
We immediately see from (\ref{tdse}) that a particularly interesting limit is to take this planar limit while keeping 
$$
\ep_2  = \hbar \quad\text{fixed}\;. 
$$
In this limit the time dependence of the differential equation (\ref{tdse}) drops out and we are left with an interesting time-independent Schr\"odinger equation for the $\ep_2$-brane.  We will devote section \ref{QsurfacesNS} to the discussion of this limit. 

On the other hand, the theory becomes classical if we in addition take $\ep_2$ to zero. In this limit, the Schr\"odinger equation becomes the Riemann surface equation \req{after},
$$-p^2 + W'(x)^2 + f(x) = 0\,.
$$ 
Here the coefficients $b_n$ are fixed in terms of the filling fractions, either by requiring 
$$\oint_{A_I} p\, dx = \mu_I = N_I\ep_1\,,$$
around the cuts that open up from the critical points of the potential $W(x)$,  or equivalently from 
the 
classical pre-potential ${\cal F}^{(0)}(\mu) $, by $b_n =  \partial_{(n)}{\cal F}^{(0)}$.  
For general $\ep_1,\ep_2$, the Riemann surface becomes quantum, as $p$ becomes an operator
$$
p = \epsilon_\alpha{\del \over \del x}\,,
$$
and moreover, the times begin to flow, as the Schr\"odinger equation is time dependent. The $b_n's$ started out as  complicated, $\epsilon_\alpha$ dependent function of the 't Hooft couplings, $\mu_I$ determined from \req{bb}.  However, since they parameterize the closed string moduli space equally well as $\mu_I$'s do (since there are as many of them as there are cuts and the B-periods), we can use them to parameterize our ignorance of the (in general) complicated closed string amplitude that underlies them. Later on, we will see that from the solutions to the Schr\"odinger equation ${ \Psi}(x; \{b_n\})$ we can in fact determine ${\Fcal_{closed}}$, at least in the NS limit.

\subsection{Virasoro Constraints and Hidden Conformal Symmetry}

The matrix model  has a conformal symmetry, which is well known in the unrefined case, and which survives refinement. It leads directly to the Schr\"odinger equation.

Just as in the usual matrix model \cite{Kostov:1999xi}, in the $\beta$-ensemble matrix model with polynomial potential $W(x)$ we can identify the scalar field as
\begin{equation}\notag
\phi(x)  = g_s^{-1}  W(x) - \frac{g_s}{\e_2} \Tr \log (x-\Phi)\,.
\end{equation}
Using the operator equation 
$$\Tr\Phi^n = -\frac{\e_2}{2} \frac{\partial}{\partial t_n}\;,
$$ 
we get the standard mode expansion of a free chiral boson
$$
\partial \phi(x)  =g_s^{-1} \sum_{k=1}^\inf k t_k x^{k-1} + \frac{g_s}{2} \sum_{k=0}^\inf x^{-k-1} \frac{\pa}{\pa t_k}\,.
$$ 

The symmetry of the matrix model corresponding to re-parametrizing the eigenvalues hides a conformal symmetry. Concretely, the Ward identity
\begin{equation}\label{ward1}
0=\sum_{i=1}^N \int  d^Nz\, \frac{\pa}{\pa z_i} \left( \, \frac{1}{x-z_i}{\cal O}\,e^{-\frac{2}{\e_2} \big(\sum_j W(z_j) + \e_1 \sum_{k< j}\log(z_k-z_j) \big)} \right) \;,
\end{equation}
ensures the invariance of physical quantities under reparametrizing the eigenvalues via
arbitrary polynomial functions.

Let $T(x)$ be the energy-momentum tensor
$$
T(x)  =\sum_{k} x^{-k-2} L_k =: \pa  \phi(x) \pa \phi(x)  : + \frac{\e_1+\e_2}{g_s}\, \pa^2 \phi(x)\, ,$$
where the second term reflects the presence of a background charge
$
Q =({\e_1+\e_2}) /{ g_s}
$
in the corresponding conformal field theory. Indeed one can show that the operators ${L}_k$ satisfy the Virasoro algebra with a central charge given by $Q$. Using the formula for $\phi(x)$ we arrive at the following expression for the energy-momentum tensor as an operator
$$
g_s^2 T(x) = 2 \ep_1 \sum_i \frac{W'(x)-X'(z_i)}{x-z_i} + W'(x)^2 + (\ep_1+\ep_2) W''(x)\;.
$$
Then one can show that the Ward identity satisfied by the partition function takes simply the form of the Virasoro constraint satisfied by the ground state wave-function
$$
{L}_n \,Z = 0  \quad,\quad n\geq -1 \;.
$$

Now we will turn our attention to the brane partition function. 
We will see that also the Ward identity equation for the brane partition function can again be written in terms of Virasoro constraints. 
But instead of the Virasoro ground state condition satisfied by the closed partition function, the brane partition function satisfies equations analogous to the BPZ equations satisfied by the degenerate states corresponding to reducible representations of the Virasoro algebra.
In particular, the two types of branes of the matrix model exactly correspond to the two types of degenerate states in the Virasoro minimal model and Liouville conformal field theory. 

The comparison with the conformal Ward identity is analogous to the vacuum equation discussed above. The only difference is, just as the energy-momentum tensor acts on the degenerate insertion as well as other operator insertions, in the matrix model energy-momentum tensor we should use the  effective potential $\til W(y)$ perturbed by the brane insertion at location $x$ and  related to the original potential $W(y)$ as
$$
\til W(y) = W(y) -\frac{\e_1\e_2}{2\epsilon_\alpha} \log(x-y) \;.
$$
This correction gives an extra term 
$$
\frac{1}{2\p i} \oint \frac{\til W'(y)^2 }{y-x} dy = W'(x)^2 -  \frac{\e_1\e_2}{\epsilon_\alpha} W''(x) $$
in the operator
$$
L_{-2}(x) =\frac{1}{2\p i} \oint dy \frac{T(y)}{y-x}\,.
$$
Using the energy-momentum operator, the differential equation \req{differential} can be written as
\be\label{differential2}
\left(-\epsilon_\alpha^{2} \pa_x^2 + g_s^2 L_{-2}\right)Z_\alpha(x) = 0
\,.
\ee
Notice that, upon a simple rewriting of the above equation as
\be\label{differential_conformal_form}
\left(b^{2} \pa_x^2 + L_{-2}\right)Z_1 = 0\,,\quad  \;\left(b^{-2} \pa_x^2 + L_{-2}\right)Z_2 = 0\,,\quad b^2 = \frac{\e_1}{\e_2}\,, 
\ee
the Ward identity equation with a brane insertion takes the form of the BPZ equation satisfied by a correlation function in two-dimensional CFT with a degenerate operator inserted at location $x$.

\section{Quantum Riemann Surface and the NS Limit}
\label{QsurfacesNS}

In section \ref{Time-Dependent Schrodinger Equation}, we have seen that an interesting limit to consider in the $\beta$-ensemble matrix model is when 
\be\label{NSlimit}
\epsilon_1 \to 0 \quad \text{with} \quad\epsilon_2=\hbar\quad\text{finite} \;.
\ee
In this limit, the $\ep_1$-brane decouples, as the corresponding coupling constant $\ep_1$ vanishes. However, the coupling constant of the $\ep_2$-brane remains finite, and the time-dependent Schr\"odinger equation \req{tdse} reduces to a time-independent Schr\"odinger equation %
\beq\label{eqt}
\Bigl(-p^2 +W'(x)^2 +f(x) \Bigr) \Psi_2(x) =0\quad,\quad \text{with}\quad[p,x] = \hbar\;.
\ee
This is the limit considered in \cite{nesh} in the gauge theory context, and we will refer to it as the NS limit.

This section will be devoted to the study of refined topological strings in the NS limit. Notice that in this limit we have $g_s^2 =- \ep_1 \ep_2 \rightarrow 0$ and we infer that the refined topological string theory becomes classical, since $g_s$ continues to play the role of the closed string coupling constant in the refined theory as we discussed in section \ref{Refined Topological Strings}. This constitutes a significant simplification. 

%

%
%
%

For the more general class of geometries introduced in section \ref{Rev}, we do not have a similarly rigorous derivation of the Schr\"odinger equation satisfied by the brane wave-function as for the (subclass of) geometries with matrix model duals. In particular, in general we do not have enough input to settle the ordering ambiguities. However it is natural to extend the methods of section \ref{MMtimedepS} to the $\beta$-deformed Toda matrix model \cite{DV09},
and show that the same result follows.  Moreover, note that by taking the rank $r$ of the Lie algebra of the Toda matrix model
arbitrarily high, we can get an arbitrarily high degree polynomial for $H(x,p)$. 
We leave the details to future work. 
Here, we will just assume that what we derived from matrix model is a general phenomenon. 
Namely, for a B-brane on any Riemann surface
$\Sigma: H(x,p)=0$ arising from a Calabi-Yau geometry, as described in section \ref{Rev}, 
%
%
%
%
%
%
we claim that in the NS limit the $\ep_2$-brane satisfies a time-independent Schr\"odinger equation with the Riemann surface as the Hamiltonian, \ie, 
\be\label{canonical_quantisation}
H(x,p) \Psi(x)=0\quad,\quad [p, x]=\hbar\;
\ee 
with $\Psi(x):=\Psi_{2}(x)$. We will test this idea in section \ref{toric} for several non-trivial toric geometries and observe that the conjecture checks extremely well, despite the absence of a matrix model and the presence of ordering ambiguities. 
A related observation that the gauge theory instanton partition function in the presence of the surface operator becomes in the NS limit the eigenfunction of the Hamiltonians of the corresponding quantum integrable systems has been previously made in \cite{Alday:2010vg,MT10,Marshakov2010,Teschner:2010je}.

In the next subsection, we show how the brane wave-functions transform under monodromy in the NS limit. This will be useful to 
elucidate the observation of \cite{nesh} relating the instanton partition function and the energies of certain quantum integrable systems, as we will explain in the following subsection.

\subsection{Brane Monodromies and the NS Partition Function}
\label{NSmonodromy}

As discussed in sections \ref{sec2shifts} and \ref{sec3shifts}, the monodromies of the exact wave-functions of the branes put constraints on the closed topological string partition function. This is true both for the usual as well as the refined topological strings. 
Combined with the Schr\"odinger equation (\ref{canonical_quantisation}) describing the probe $\ep_2$-brane wave-function, this gives important and computable information about the closed string partition function in the NS limit. 

Solving the Schr\"odinger equation for $\Psi_{2}(x)$ and writing
$$
\Psi_{2}(x) = \langle e^{-\frac{1}{\ep_2}\phi(x)} \rangle = e^{ \frac{1}{\ep_2}\int^x \partial S}\;,
$$
we conclude that taking the brane around $\gamma_B= \sum_I m_I B^I$ induces a change in the phase of the {\it closed} string partition function by
\beq\label{br}
Z_{\text{\it top}}(\vec a) \rightarrow e^{\frac{1}{\ep_2} \oint_{\gamma_B} \del S } Z_{\text{\it top}}(\vec a)\,.
\ee
On the other hand, as we have seen in section \ref{sec3shifts} on general grounds, taking an $\ep_2$-brane
around $\gamma_B$ changes the closed string partition function by
$$
{\cal M}_{\gamma_B} :\;Z_{\text{\it top}}(\vec a) \rightarrow  Z_{\text{\it top}}(\vec a + \tfrac{g_s^2}{\epsilon_2} \vec m)\,.
$$
On the other hand, we know that the following is true in the NS limit we have $g_s \to 0$ and hence
$$
Z_{\text{\it top}}(\vec a; \ep_1,\ep_2)=\exp\left(\frac{1}{g_s^2}{\cal F}^{(0)}(\vec a;\hbar)+ \dots \right) \;,$$
where the dotted terms are suppressed by a factor of $g_s^2$. 
On the other hand, the finite shift due to the brane monodromy becomes infinitesimal, so $Z_{\text{\it top}}(\vec a)$ changes by
\beq\label{dt}
\lim_{\substack{\ep_1\to 0, \\ \ep_2 = \hbar}}\,{\cal M}_{\gamma_B} :\;Z_{\text{\it top}}(\vec a) \rightarrow \exp\left(\frac{1}{\hbar}\sum_I m_I \del_{a_I} {\cal F}^{(0)}(\vec a;\hbar)\right)Z_{\text{\it top}}(\vec a)\,.
\ee

Consistency of the two equations \req{br} and \req{dt} implies that as we take the $\ep_2$-brane around a $B^I$ cycle, its phase changes by 
\beq\label{bcycle}
\oint_{B^I} \del S  =   \del_{a_I} {\cal F}^{(0)}(\vec a;\hbar)\,.
\ee
Note that this is accompanied by
\beq\label{acycle}
\oint_{A_I} \del S  =  a_I(\hbar)\,,
\ee
by definition of what we mean by the $A_I$-cycle. Solving for the brane wave-function explicitly from (\ref{canonical_quantisation})
given the classical curve equation $\Sigma: H(x,p)=0$ (involving some arbitrary coordinates on the Calabi-Yau moduli space) 
and computing the periods \req{bcycle} and \req{acycle} from brane monodromies around the corresponding cycles hence defines a `quantum' special geometry on the Calabi-Yau moduli space. This is an $\hbar$-dependent generalization of the usual special geometry. In particular, the $a_I(\hbar)$ periods obtained in this way are good, `flat' coordinates on the moduli space.

\subsection{An Explanation of the Results of Nekrasov-Shatashvili}

From the point of view of four-dimensional gauge theory, the NS limit (\ref{NSlimit})
is clearly an interesting one. In this limit, two of the four dimensions where the Poincar\'e invariance is broken become effectively compactified and we are left with a system that is effectively two-dimensional. Indeed, in \cite{nesh} Nekrasov and Shatashvili observed that the gauge theory in the  $\Omega$-background in this limit is connected to various quantum integrable systems in a highly non-trivial way, a feature shared by various two-dimensional supersymmetric gauge systems \cite{Nekrasov:2009zz,Nekrasov:2009uh,Nekrasov:2009ui,Gerasimov:2006zt,Gerasimov:2007ap,Moore:1997dj}. 
The Bethe ansatz equation for the corresponding integrable system turns out to coincide, according to \cite{nesh},
with the critical points of the free energy in this limit.  More precisely it is given by
\be\label{bethe}
\exp\left( \frac{\partial {\cal W}(\vec a;\hbar)}{\partial a_I}\right) =1\,.
\ee
These equations determine $a_I$ given a $g$-tuple of integers $n^I$ by
$$\frac{\partial {\cal W}(\vec a;\hbar)}{\partial a_I}=2\pi \ii n^I\,.$$
The function ${\cal W}$ is related to the refined gauge theory partition function $Z_{{\cal N}=2}(\vec a; \ep_1,\ep_2)$ at arbitrary $\ep_{1,2}$ via 
$$
{\cal W} = \lim_{\ep_1 \rightarrow 0} \ep_1 \log Z_{{\cal N}=2} \,.
$$
Furthermore, the corresponding eigenvalues of the $g$ commuting Hamiltonians can be expressed
in terms of functions of the solutions $a_I$ to the above equation.  The question to address is why
studying the partition functions of a 4d gauge theory system and taking such a limit should have anything
to do with answering these questions for an integrable model.  Here we would like to explain this fact.
The basic point will be the following:  We can study an ${\cal N}=2$ gauge theory system in 4d in two
different ways. The first one uses a target space description while the second one uses world-sheet techniques.  The fact that  these
two approaches have to render the same results will be the key to an explanation of the observation in \cite{nesh}. A related consideration has also
appeared in \cite{Dorey:2011pa}. See also \cite{Chen:2011sj}.

In particular, the NS computation is the target space viewpoint, and the integrable system emerges
from the world-sheet viewpoint. The equality of $Z_{{\cal N}=2}$ and the refined topological string partition function $Z_{\text{\it top}}$ implies
$$
{\cal W}(\vec a, \hbar) = \tfrac{1}{\hbar}{\cal F}^{(0)}(\vec a;\hbar) \,,
$$
where 
${\cal F}^{(0)}(\vec a; \hbar)$ is the genus zero refined topological string amplitude.

Consider an ${\cal N}=2$ system given by a Seiberg-Witten curve
$\Sigma: H(x,p)=0$. 
We assume, as in the rest of this paper, that the corresponding SW differential
is $\lambda=p\,dx$.  This theory can be engineered in type IIB strings by considering
a local CY given by a hypersurface with the equation (\ref{hypersurface}),
or in M-theory in terms of M5 brane wrapping the curve $\Sigma$ in the complex
two dimensional space $(x,p)\subset {\C}^2$.
Let us assume this curve has genus $g$.  Then we know that there are $g$ deformations of this curve
given by the $g$ Coulomb parameters $a_{I=1,\dotsi,g}$.  Let us further assume, as is the case in all the known examples,
that the $a_I$ dependence of $H$ can be written as
$$H(\vec a; x,p)=H_0(x,p)+\sum_{I=1}^g f_I(x,p) E_I(\vec a)$$
for some parameters $E_I$ that are functions of the flat coordinates $a_I$.
The NS prescription turns out to translate in this language to the
statement that $a_I$ are fixed by the `critical points' of ${\cal W}$ and these specify the eigenvalues $E_I$
of the $g$ commuting Hamiltonians. 

The key idea to connect the integrable system to the refined topological string considerations is a trick known as the
`separation of variables' (SOV) \cite{Sklyanin:1995bm}.  Think about ${\C}^2$ as a complexified classical
phase space.  Consider $g$ points on it given by $(p_I,x_I)$, as $I$ runs from $1,...,g$.   We can choose
the moduli of the SW curve to pass through all these points.  In particular this fixes all the $E_I$ in the above
equation in terms of these $g$ points: 
$$E_I = h_I(p_1,x_1;p_2,x_2,...)\,,$$
for some functions $h_I$ that are readily computed from requiring that
$$
H(p_J,x_J) =0\quad, \quad J=1, \ldots g\,.
$$
It is not difficult to see that the defined $h_I$ are classically, as well as quantum mechanically, commuting and thus define
an integrable structure (\ie, we have $g$ commuting Hamiltonians $h_I$ for a $2g$ dimensional phase space $(p_I,x_I)$) \cite{Babelon:2002wt}.  Furthermore, it can be shown that in the quantum theory the eigenstates of $h_I$ can be written using the separation
of variables in terms of single-particle wave-functions:
$$\Psi (x_1,...,x_g)=\prod_I\Psi_I(x_I)\,,$$
where each $\Psi_I(x_I)$ is annihilated by $H(x,p)$. Namely, they separately satisfy the equation (\ref{canonical_quantisation}). 
%
%
In the context of refined topological strings, we identify the corresponding wave-functions
$\Psi(x)$ as the amplitudes for the branes in the refined open topological strings, in the NS limit.
We are almost finished except for the fact that  the brane wave-functions are holomorphic functions defined on the complex
$x$-plane and not just on a real space. 
Hence, the wave-functions could have monodromies and the question of their single-valuedness arises. 
To study this issue, let us assume that we have chosen
a real subspace as well as a choice of
the real coefficients $h_I$ in the SW geometry, such that $H=0$ leads classically to $g$ physically
allowed intervals in the $x$ space.  Let us call these intervals $b^I$, for $I=1,...,g$. Note that
the end points of the classical intervals correspond to $p=0$ and in the hyperelliptic case, and correspond
to branch points of the SW curve over the $x$-plane.
Let us assume
that the exact wave-functions have $n^I$ zeros in the interval $b^I$.    Consider cycles $B^I$  in the complex plane
encircling $b^I$.  These would correspond to a basis of B-cycles of the SW curve.  The fact
that $\Psi$ has $n^I$ zeroes on the interval $b^I$ implies that in the analytically continued picture, surrounding the $B^I$ cycle picks up a phase $\exp(2\pi i n^I)$.  This would also guarantee that the
restriction of $\Psi$ to the real slice is single-valued and well-defined.
On the other hand, we have seen in the last
section in (\ref{dt}) that taking the
brane around the $B^I$ cycles changes its phase by the derivative of the closed string partition function $\partial_I{\cal W}(\vec a)$:
$$\Psi(x+B^I)=\exp\left( \frac{\partial {\cal W}(\vec a;\hbar)}{\partial a_I}\right) \Psi (x)\,.$$
Putting the two together, we conclude that
$$\frac{\partial {\cal W}(\vec a;\hbar)}{\partial a_I}=2\pi i n^I\,,$$
thus explaining the results of NS.
This relation between the phase of the wave-function and the integers $n^I$ is known as the `exact
Bohr-Sommerfeld' quantization rule.  It generalizes the usual rule to a form which is true to all orders
in $\hbar$.  In this context, the relevance of the NS limit was also pointed out in \cite{mm1,mm2,Marshakov2010}.

\paragraph{The scaling $n^I\sim \beta \tilde N^I$}
The quantization of $n^I$ for the solutions of the Bethe-ansatz equations is reminiscent of
the filling fractions in the context of matrix models.  Let us consider the cases where the topological string theory is dual
to a matrix model. Then consider a closely related matrix model with a number of critical points, for which the above physical intervals $b^I$ become the cut around the critical points in the eigenvalue plane ${\mathbb C}$ at large $N$. By considering this other matrix model we are effectively exchanging what we call the A- and the B-cycle. Denote by  $\tilde N^I$ the filling fractions for each of the critical points.
As noted before, the NS limit corresponds to 
taking $\epsilon_1\rightarrow 0, \tilde N^I\rightarrow \infty$ while keeping the 't Hooft coupling $\tilde \mu^I = \epsilon_1\tilde N^I$ finite.  And the period around the cycle $B^I$ encircling the cut is given by
$$
\oint_{B^I} p\,dx =\tilde N^I \epsilon_1\,.
$$
On the other hand, in the WKB approximation the Bethe-ansatz equation reduces to
$$\oint_{B^I} p\,dx=n^I \epsilon_2\,.$$
Thus in this limit we have
$$n^I\epsilon_2=\tilde N^I\epsilon_1\Leftrightarrow n^I=\tilde N^I\beta\,.$$
Of course this relation is only valid to leading order in $\hbar=\epsilon_2$.

\section{Genus Zero Examples}
\label{MMexamples}

In the context of matrix models we have shown that if we consider the Gaussian matrix model
the brane wave-functions satisfy the exact equations
$
H(x,p) \Psi_\alpha(x) = 0
$
with
$
[p,x] = \epsilon_\alpha\;,
$
where 
$$
H(x,p)=-p^2+x^2-\mu\;.
$$
In particular the time-dependent Schr\"odinger equation \req{tdse} has trivial time dependence that has been
absorbed in the shift of the background, and therefore we obtain the same equation in the NS limit
as in the general case without taking the NS limit.   For multi-cut matrix models with have higher
degree potentials, this is no longer true.  In these cases the genus of the spectral curve is greater
than zero, and the multi-time dependence of the Schr\"odinger equation captures the
moduli of the higher genus curve, which are absent for the genus zero case.  It is thus natural to conjecture
that for all the cases where
$\Sigma: H(x,p)=0$
is a genus zero curve, as in the Gaussian matrix model example, the time dependence of the Schr\"odinger equation is absent.
As already noted, the exact equation satisfied by the branes for any $\ep_1,\ep_2$ is the time independent Schr\"odinger equation. 

In more details, on a genus zero Riemann surface, in general we have a set of compact A-cycles. All the B-cycles are by necessity non-compact, though we can find a basis of them satisfying $A_I\cap B^J = \delta_{I}^J$.
The monodromies around the A-cycles define the flat coordinates
$$
\oint_{A_I} \del S/\ep_1 = a_I/\ep_1\;.
$$

In particular, the partition function of the brane depends only on one $\epsilon_\alpha,$ associated to the corresponding brane
$$\Psi_\alpha(x) = \exp(S(x,\epsilon_\alpha)/\epsilon_\alpha)\;.
$$
Solving the Schr\"odinger equation, we can find not only the brane partition functions, but also the closed string partition functions. In general, on a genus zero Riemann surface we have a set of compact A-cycles. All the B-cycles are by necessity non-compact, though we can find a basis of them satisfying $A_I\cap B^J = \delta_{I}^J$.
The monodromies around the A-cycles define the flat coordinates
$$
\oint_{A_I} \del S/\ep_1 = a_I/\ep_1\;.
$$
Consider now taking the brane $\Psi_{1}$ brane around a $B$-cycle. On the one hand, from general considerations
$$
\oint_{B_I} \del S/\ep_1  = {\cal F}(a_I+g_s^2/2\ep_1)/g_s^2 - {\cal F}(a_I-g_s^2/2\ep_1)/g_s^2\;,
$$
where ${\cal F}(\vec a; \ep_1,\ep_2)=g_s^2\log Z_{\text{\it top}} $ is the exact free energy of the refined topological string.
On the other hand, since the solution for the brane is independent of $\ep_2$, and taking the NS limit $\ep_2 \rightarrow 0,$
we find that the same amplitude equals $\del_I {\cal W}$: 
$$
\oint_{B_I} \del S/\ep_1  = \del_I{\cal W}\;.
$$ 
Hence, we derive that 
$${\cal F}(a_I+g_s^2/2\ep_1)/g_s^2 - {\cal F}(a_I-g_s^2/2\ep_1)/g_s^2 = \del_I {\cal W}\,,$$
and both equal to the change of phase of the brane, after being transported around the B-cycle. 
The fact that the two expressions agree is special for genus zero Riemann surfaces. In general, monodromies of the branes give rise to difference equations for the partition functions. In the NS limit, the difference equation becomes a differential equation, for ${\cal W}$.

In this section we present two classes of examples along these lines.  The first one involves the Gaussian
matrix model and the second class involves toric geometries without compact 4-cycles.

\subsection{The Gaussian}
\label{The Gaussian}
The simplest matrix model to which our derivation of the Schr\"odinger equation of section \ref{MMtimedepS} applies is the Gaussian
$$
Z = \int d^N z \prod_{i<j} (z_i - z_j)^{-2\ep_1/\ep_2}  e^{- \frac{1}{2\ep_2} \sum_k z_k^2}\,.
$$
The classical spectral curve is the deformed conifold
\beq\eqlabel{GaussianH}
H(x,p)= -p^2+x^2-\mu=0\,,
\eq
which has one A-cycle, corresponding to the circle at real values of $x$ and $p$, and one non-compact B-cycle. The $\beta$-ensemble matrix model partition function can be easily evaluated explicitly to find the free energy (under the shift $\mu\rightarrow\mu+(\ep_1+\ep_2)/2$)
\beq\notag
{\cal F}(\mu)/{g_s^2}= \log Z =\int \frac{ds}{s}\frac{e^{-\mu s}}{(e^{\ep_1s/2}-e^{-\ep_1s/2})(e^{\ep_2s/2}-e^{-\ep_2s/2})}\;.
\ee
One infers that this is exactly the partition function of the $c=1$ string at radius $R=1/\beta$ \cite{DV09}, which has been computed in \cite{GK90}.  Note that the NS limit in this case corresponds to the $R\rightarrow \infty$ limit at fixed $\mu$.  In particular, the partition function {\it per unit volume} corresponds precisely
to the NS definition of the partition function in this limit:
$${\cal W}(\mu)=\lim_{\ep_1\rightarrow 0 } \ep_1 \log Z\;,$$
with
\beq\notag
\del_{\mu} {\cal W} =  \int \frac{ds}{s}\frac{e^{-\mu s}}{(e^{\ep_2s/2}-e^{-\ep_2s/2})}\;.
\ee
Note that from this it follows immediately that
$$
\Bigl({\cal F}(\mu + g_s^2/2\ep_1) - {\cal F}(\mu -g_s^2/2\ep_1)\Bigr)/g_s^2 = \del_{\mu} {\cal W}
$$
in agreement with our claim. 
We want now to check that we can indeed recover in the NS limit the coefficients \req{GaussianNSlimit} via the monodromies of the brane solving the Schr\"odinger equation we derived in sections \ref{MMtimedepS} and \ref{QsurfacesNS}. For this, we will need the power series expansion of ${\cal W }$:
$$
{\cal W} ={1\over 2}\mu^2 \log\mu+ {\cal W}^{(0)}\log\mu+\sum_{n>0}\frac{\ep_2^n}{\mu^n}{\cal W}^{(n)}\,,
$$
where
%
\beq\eqlabel{GaussianNSlimit}
{\cal W}^{(0)}=\frac{1}{24}, \qquad
{\cal W}^{(n)}=\frac{(2^{-n-1}-1)}{n(n+1)(n+2)}B_{n+2}\,.
\eq


Adding an $\ep_1$-brane, the Riemann surface becomes quantum:
$$
(-\ep_2^2 \partial_x^2 + x^2) \Psi(x) =  \mu \Psi(x)\;.
$$  
 
This is the (inverted) quantum harmonic oscillator.  
The monodromies of the wave-function $\Psi(x)$ around the B-cycle have been in fact calculated to all orders in $\hbar$ already some time ago in the context of Stokes phenomena in quantum mechanics, see for instance \cite{voros}. (We let $\hbar=\ep_2$ for the rest of this section.) The result can be matched with \req{GaussianNSlimit}. Alternatively, they can also be computed using the matrix model techniques of solving the loop equations. This method has been used to solve the $\beta$-ensemble Gaussian matrix model in \cite{Mironov:2011jn}.
%
%
%
%

%
%

In order to illustrate the technique we will use later in the context of the more complicated cubic potential in section \ref{toric}, let us re-derive this result in a simpler, but non-exact fashion. In detail, we can solve the Schr\"odinger equation for the brane wave-function $\Psi(x)$ via the usual WKB Ansatz known from elementary quantum mechanics, \ie,\beq\eqlabel{WKB1}
\Psi(x)=\exp\left(\frac{1}{\hbar}S(x)\right)\,,
\eq
with
\beq\eqlabel{WKB2}
S(x)=S_0(x)+\sum_{n=1}^\infty S_n(x)\,\hbar^n\,.
\eq
Plugging this Ansatz into the Hamiltonian and expanding in $\hbar$ yields for the first few orders
\beq\eqlabel{Snp}
\del_xS_0=p\,,\,\,\,\,\del_xS_1=-\frac{1}{2} \partial_x\log p\,,\,\,\,\, \del_xS_2=-\frac{(\partial_x\log p)^2-2 \partial^2_x\log p}{8 p}\,,\dots\,.
\eq
Note that the WKB approximation determines directly not $S(x)$, but $\del_x S$. Near any point on the Riemann surface, we can integrate this to obtain $S(x)$ locally, however the global solutions have monodromies.
In particular, to obtain from the semi-classical wave-function $\Psi(x)$ the (quantum corrected) periods $\Pi_{\Ccal}=\sum_{n=0}^\infty \Pi_{\Ccal}^{(n)}\,\hbar^n$, we have to integrate $\del S$ along the classical cycles of the geometry. 

For specific examples, or better potentials $V(x)$, things considerably simplify, since we can turn the contour integration for the higher order corrections into a differentiation. For example, this idea has been put forward in the gauge theory context in \cite{mm1}.  In detail, we can argue for differential operators $\Dcal^{(n)}$ such that
\beq\eqlabel{DopDef}
\Pi_\Ccal^{(n)}=\Dcal^{(n)}\, \Pi^{(0)}_\Ccal\,.
\eq
Defining
\beq\notag
\Dcal=1+\sum_{n=1}^\infty \Dcal^{(n)}\hbar^n\,,
\eq
the full quantum corrected periods are simply given by $\Pi_\Ccal=\Dcal \Pi_\Ccal^{(0)}$.
\footnote{
Note that we know from the M-theory picture described in section \ref{refMtheory} that the refined partition function in the NS limit is odd in $\hbar$, hence ${\cal D}^{(n)}$ with $n$ odd vanishes. In particular, since $S_n'$ is always a total derivative for $n$ odd, this implies the integral of total derivatives vanish and there is no subtleties involving singularities. }

We can partially integrate the contour integrals of \req{Snp} to obtain the more convenient expressions \footnote{For simplicity, we absorbed a normalization factor of $\sqrt{2}$.}
\beq\eqlabel{partialS}
\begin{split}
S'_0(x)&=\sqrt{(V(x)-E)}\,,\\
S'_2(x)&=\frac{V''(x)}{48(V(x)-E)^{3/2}}\,,\\
S'_4(x)&=-\frac{7V''(x)^2}{1536(V(x)-E)^{7/2}}-\frac{V''''(x)}{768(V(x)-E)^{5/2}}\,,\\
&\vdots
\end{split}
\eq
where we used the parametrization $p(x)=\sqrt{W'(x)^2+f(x)}=:\sqrt{V(x)-E}$. For the Gaussian, we can identify $V(x)=x^2$ and $E=\mu$ and the operators $\Dcal^{(n)}$ are particularly simple because $V''(x)=2$ and $\partial_x^{>2} V(x)=0$. Namely, one has
\beq\eqlabel{ConievenDops}
\Dcal^{(2)}=-\frac{1}{24}\partial_\mu^2\,,\,\,\,\,\, \Dcal^{(4)}=+\frac{7}{5760}\partial_\mu^4\,,\,\,\,\,\,\Dcal^{(6)}=-\frac{31}{967680}\partial_\mu^6\,,~\dots\,.
\eq
We conclude that the period $\Pi_A^{(0)}$ does not receive any quantum corrections, while we obtain for the B-period $\Pi_B=\Dcal\Pi_B^{(0)}$,

\beq\eqlabel{defconiFh}
\Pi_B(\mu)=\Pi_B^{(0)}(\mu)-\frac{1}{24}\mu^{-1} \hbar^2+\frac{7}{2880}\mu^{-3}\hbar^4-\frac{31}{40320}\mu^{-5}\hbar^6+\dots\,,
\eq
or, after integrating over $\mu$,
\beq\eqlabel{defconiWh}
\hbar\,\Wcal(\mu)=\Fcal^{(0)}(\mu)-\frac{\hbar^2}{24}\log\mu-\frac{7}{5760}\mu^{-2}\hbar^4+\frac{31}{161280}\mu^{-4}\hbar^6+\dots\,,
\eq
in agreement with the expectation \req{GaussianNSlimit}.

\subsection{Toric Geometries without 4-cycles}

Consider the A-model topological string on a toric  Calabi-Yau $X$ without compact $4$-cycles. 
This is mirror to the B-model on the Calabi-Yau 
$$
uv + H(e^{x}, e^p) = 0\;.
$$
The corresponding Riemann surface has genus zero and is of the form
\beq\label{tor1}
H(e^{x},e^{p}) = P_n( e^{x}) + e^{p+ m x} P_k( e^{x})=0\;,
\ee
where $P_{n,k}(e^{x})$ are polynomials of degree $n$ and $k$ in variable $e^{x}$, and $m$ is an arbitrary integer.
This Calabi-Yau has $n+k-1$ moduli which enter as coefficients of the polynomials.  
The simplest example is the conifold with $n=k=1$ and $m$ arbitrary.
As explained in \cite{av00,akv01}, there is a particularly natural way of writing the Calabi-Yau geometry in terms of open and closed string flat coordinates that are typically given by certain periods on the mirror geometry.
By a change of variables we can rewrite \req{tor1} as
\beq\label{flat}
H( e^{\hat x}, e^{\hat p} ) = \prod_{I=0}^n(1- Q_{\alpha_I} e^{\hat x}) - e^{\hat{p}+ m \hat{x} }\prod_{J=0}^k(1- Q_{\beta_J} e^{{\hat x}})\;.
\ee
Depending on the chamber, it is natural to set one of $Q_{\alpha_I}$ or $Q_{\beta_J}$ to $1$, which we can do by a leftover degree of freedom to shift ${\hat x}$ and  $\hat{p}$ by constant values. This gives indeed altogether $k+n-1$ moduli. The Riemann surface can be viewed as a copy of a ${\hat u}$ cylinder, where the one-form
$$
\del \phi = \hat{p}\,d {\hat x}
$$
has singularities at $\hat{x} ={ t}_{\alpha_I} $, $\hat x ={t}_{\beta_J}$. Up to a gauge degree of freedom discussed above, 
${\hat x}$, $t_{\alpha_I}= -\log{Q_{\alpha_I}} $ and $t_{\beta_J}=- \log Q_{\beta_J}$ are flat coordinates on the open and closed moduli space. The canonically conjugate variables are
$$
[\hat p,\hat x] = \ep_1
$$
In particular, the brane $\Psi(\hat x)$ satisfies the exact Schr\"odinger equation
\beq\label{tse}
H(e^{\hat x},e^{\hat p} )\Psi(\hat x) =0\;,
\ee
which is a difference equation, 
$$
\prod_{I=0}^n(1- Q_{\alpha_I} e^{\hat x}q_1^{-1/2}) \Psi(\hat x)- e^{m \hat{x} }\prod_{J=0}^k(1- Q_{\beta_J} e^{{\hat x}}q_1^{-1/2})\Psi(\hat x+\hbar) =0\;,
$$
where $q_1 = e^{-\ep_1}$. Note that we have redefined the K\"ahler parameters $t_\alpha$ and $t_\beta$ with a shift.  
This has exact solutions in terms of quantum dilogarithm functions  
%
$$
\Psi(\hat x)=e^{-\frac{m {\hat x}^2}{2\hbar} +\frac{m \hat x}{2}}\frac{ \prod_{I=0}^n L(Q_{\alpha_I} e^{\hat x}) }{\prod_{J=0}^kL(Q_{\beta_J} e^{{\hat x}})},
$$
where
$$L(e^x) = \prod_{\ell = 1}^{\infty} (1-q_1^{\ell-1/2}e^x)\;. 
$$
Note that $L(e^x)$, given by the quantum dilogarithm, satisfies
$$L(e^{x+\ep_1}) = (1-q_1^{-1/2} e^x) L(e^x)\;.$$
This includes, for example, the conifold, where 
%
$$
\Psi_{\text{\it conifold}}(\hat x)=\frac{L(e^{\hat x}Q)}{L(e^{\hat x})}
$$
solves \req{tse} with
$$H_{conifold}(\hat x, \hat p) = (1-e^{\hat x}q_1^{-1/2})e^{\hat p} -(1-e^{\hat x} Qq_1^{-1/2}) \;.
$$
Note that in general there are ordering ambiguities in defining the Hamiltonians, but they can all be absorbed into the shifts of the open and closed moduli, \ie, what we mean by ${\hat x}$, $t$. As a consequence, in this particular case, there is really no physics behind the quantum shifts, and we can simply choose them in the way that is the most convenient. Any two such choices differ by re-parametrization of the moduli space, and hence are physically equivalent.
Finally, the above expressions all assume we are in the regime of the moduli space where $Qe^{\hat x}<1$. If that fails to be the case, for example due to moving the brane around the Riemann surface, it is better to use the analytic continuation
%
%
$$L(e^{x})  = e^{-({ x}^2 + i\pi x )/\ep_1}L(e^{-x})
$$
to rewrite the wave-function.

Consider now moving the branes around. First, it is easy to show, in these coordinates the variables $ t$ and  $\hat x$ coordinates we defined via \req{flat} are already the open and closed A-periods. Secondly, since there are no compact B-periods, the branes are essentially single-valued on the Riemann surface (the only shifts come from the classical piece of the wave-function). As a consequence, computing the B-periods amounts to evaluating $\Psi$ at the poles of the Riemann surface. The good B-periods are obvious either from the A-model -- they are mirror to non-compact 4-cycles, or equivalently, by asking that the only infinities in evaluating the periods come from the classical pieces of the wave-function, corresponding to the fact that at infinity, all instanton corrections are suppressed. Consider, for example, the B-period corresponding to bringing a brane in from infinity to a singularity with ${\hat x} = {t_{\alpha_I}}$, and sending the brane back out to infinity at the singularity with $\hat x \rightarrow \infty$. Correspondingly, we find
$$
\exp\left(\frac{1}{\ep_1}\int_{B^{I}} \del S \right)= \frac{\Psi(t_{\alpha_I})}{\Psi(\infty)}
$$
up to the contribution of the classical pieces $x^2/\ep_1$ to the answer -- these are ambiguous and reflect the non-compactness of the geometry. On the other hand,  this non-compact B-cycle intersects all the 2-cycles whose areas are of the form 
$$\int_A k =  \sum_I n^I t_{\alpha_I}$$ by
$$B^{I}  \cap  A = n^I.$$
This implies that the free energy jumps by
$$
\exp\left(\frac{1}{\ep_1}\int_{B^{I}} \del S \right)=\frac{Z(t_{\alpha_I}+\tfrac{\ep_2}{2} )}{Z(t_{\alpha_I}- \tfrac{\ep_2}{2})}
$$
or
$$
\frac{ \Psi(t_{\alpha_I})}{\Psi(\infty)}=\frac{Z(t_{\alpha_I}+ \tfrac{\ep_2}{2})}{Z(t_{\alpha_I}- \tfrac{\ep_2}{2})} \;. 
$$
To compute $\Psi(t_{\alpha_I})$ we need to analytically continue all the terms in the solution above corresponding to which $Qe^{\hat x}$ becomes greater than one at $\hat x = t_{\alpha_I}$.  There are similar equations with $t_{\alpha}$ replaced by $t_{\beta}$.  We can view this as $n+k$ equations for the partition function $Z(t_{\alpha}, t_{\beta})$ depending on $n+k$ parameters, which we can solve, with the boundary condition, that in the limit where all the $t$'s we recover the partition function of $n+k$ disconnected copies of ${\mathbb C}^3$.
It is easy to see that the solution to these equations is the refined partition function of the closed topological string on this Calabi-Yau 
%
%
$$
Z(t_{\alpha_I}, t_{\beta_J})= M(1)^{n+k}
\frac{\prod_{0\leq I<I'\leq n}M(Q_{\alpha_I \alpha_{I'}})\prod_{0\leq J<J'\leq k}M( Q_{\beta_J\beta_{J'}})}{ \prod_{I, J=0}^{n,k}M(Q_{\alpha_{a}\beta_J}) },
$$
%
%
where, {\it e.g}, $Q_{\alpha_I\beta_J}$ is understood to be $Q_{\alpha_I}/Q_{\beta_J}$ if ${\text{Re}}(t_{\alpha_I})> {\rm{Re}}( t_{\beta_J})$ and
$Q_{\beta_J}/Q_{\alpha_I}$ otherwise, and similarly for the rest,
and
$$
M(Q) =  \prod_{\ell_{1}, \ell_{2} = 1}^{\infty} (1-q_1^{\ell_1-1/2} q_2^{\ell_2-1/2} Q)^{-1} \;$$
is the refined MacMahon function, where we have defined $q_2 = e^{\ep_2}
$. In the above we have repeatedly used relations such as 
$
L(Q)=\frac{M(Qq_{2}^{1/2})}{M(Qq_2^{-1/2})} .
$
For example, for the conifold, we get simply
$$
Z_{\text{\it conifold}}(t )= \frac{M^2(1)}{M(Q)}\;.
$$
%

The fact that this is also consistent with the monodromies being computed by ${\cal W}$ can be seen as follows.
Note that we can write the free energy of the topological string, ${\cal F}/g_s^2  = \log(Z)$ as
$$
{\cal F}/g_s^2 = \sum_{0\leq I<I'\leq n}\gamma(t_{\alpha_I \alpha_{I'}})+ \sum_{0\leq J<J'\leq k}\gamma( t_{\beta_J\beta_{J'}}) -
\sum_{I, J=0}^{n,k}\gamma(t_{\alpha_{I}\beta_J}) ,
$$
where
\beq\eqlabel{Fconi}
\gamma(t;\ep_1,\ep_2) = \log M(Q)\,.
\ee
In particular,
$$\gamma(t, \ep_1,\ep_2) =\sum_{n=1}^\infty \frac{Q^n}{n[n]_{\ep_1}[n]_{\ep_2}}\;,
$$
with $[n]_{\epsilon_\alpha}:=(q_\alpha^{n/2} - q_\alpha^{-n/2})$, and $Q = e^{-t}$.
On the other hand, the reduced partition function 
${\cal W} =\lim_{\ep_2 \rightarrow 0} {\cal F}/\ep_1$
is built out of the function
$$\gamma_{NS}(t, \ep_2) =  \sum_{n=1}^\infty \frac{Q^n}{n^2[n]_{\ep_2}}$$
as
$$
{\cal W} = \sum_{0\leq I<I'\leq n}\gamma_{NS}(t_{\alpha_I \alpha_{I'}})+ \sum_{0\leq J<J'\leq k}\gamma_{NS}( t_{\beta_J\beta_{J'}}) - \sum_{I, J=0}^{n,k}\gamma_{NS}(t_{\alpha_{I}\beta_J}) \;.
$$
The fact that
$$\gamma(t+\tfrac{\ep_2}{2}) - \gamma(t-\tfrac{\ep_2}{2}) = \del_t \gamma_{NS}
$$
immediately implies that we could have equivalently written
$$
\exp\left(\int_{B^{I}} \del S/\ep_2\right)=\exp(\del_{t_{\alpha_I}} {\cal W}),
$$
in agreement with our claim.

\section{Genus One Examples}
\label{toric}
Let us now consider some more non-trivial geometries and confirm that also for them  the NS limit of the refined topological string partition function can be recovered from a brane probing the quantum geometry, as expected from the discussion of section \ref{QsurfacesNS}.


\subsection{The Cubic}
Let us consider the matrix model with cubic potential
\beq\notag
W(x)=g\left(\frac{1}{3}x^3+\frac{\delta}{2}x^2\right)\,.
\eq
The $\beta$-ensemble matrix model partition function can be calculated perturbatively, as for example done in \cite{MS1}. The most leading $\hbar$-correction to the free energy of the cubic $\beta$-ensemble matrix model has also been computed in \cite{Brini:2010fc} using the loop equation techniques. Here we will use the techniques discussed in section \ref{QsurfacesNS} and obtain more non-trivial higher order results. 


As a check of the statements of section \ref{NSmonodromy}, we now want to compute the first few terms in the $\hbar$-expansion in the refined  topological string free energy for the cubic by studying the brane wave-function satisfying the Schr\"odinger equation of section \ref{QsurfacesNS}, which can be obtained by simply canonically quantizing the spectral curve. For that, note that in the case of the cubic, the 1-form of the matrix model spectral curve \req{HDV} can be expressed in terms of the branch-points (roots) $x_{1,\dots,4}$ as   
\beq\eqlabel{DVwroot}
\lambda=dx~g \sqrt{(x-x_1)(x-x_2)(x-x_3)(x-x_4)}\,.
\eq 
Following \cite{civ01}, we define new variables 
\beq\eqlabel{DVcubicvars}
\begin{split}
z_1=\frac{1}{4}(x_4-x_3)^2\,,&\,\,\,\,\,z_2=\frac{1}{4}(x_2-x_1)^2\,,\\
Q=\frac{1}{2}(x_1+x_2+x_3+x_4)\,,&\,\,\,\,\, {\cal I}=\frac{1}{2}(x_3+x_4-x_1-x_2)\,.
\end{split}
\eq
In terms of these variables, explicit expressions for the periods $\Pi^{(0)}_{A_1},\Pi^{(0)}_{A_2},\Pi^{(0)}_{B_1}$ and $\Pi^{(0)}_{B_2}$ can be easily found via expansion for small $z_1$ and $z_2$ of the corresponding contour integrals of $\lambda$  (see appendix B of \cite{civ01}). As usual, we will set $Q=-\delta$ and ${\cal I}=\sqrt{\delta^2-2(z_1+z_2)}$.

On the other hand, it is known that the B-model refined 1-loop amplitude on this geometry is given by
\beq\eqlabel{DV1loop}
\Fcal^{(1)}(s_1,s_2;\beta)=\frac{1}{2}\log \det g+ \sum_{I=1}^3\kappa_I(\beta)\log(\Delta_I)\,,
\eq
where $s_{1,2}$ denote the flat coordinates given by the A-periods, $z_{1,2}$ are other coordinates given in (\ref{DVcubicvars}), $g_{ij}=\partial_{s_i} z_j$,  $\Delta_I$ are the three discriminant loci (\cf, \cite{kmt})
\beq\notag
\Delta_1=z_1z_2\,,\,\,\,\,\Delta_2=\sqrt{\delta^2 - 2 (z_1+z_2)} \,,\,\,\,\,\Delta_3=(\delta^2-3z_1)^2-6z_2\delta^2+9 z_2^2+14z_1z_2\,,
\eq
where $\kappa_I(\beta)$ are $\beta$-dependent coefficients. 
In particular, we expect that in the B-model all the information about the refinement is encoded in the coefficients $\kappa_I(\beta)$, hence is of a purely holomorphic nature. 

It remains to fix $\kappa_I(\beta)$ by the behavior of the topological strings amplitude near the singular points. 
Since $\Delta_1$ is a conifold like singularity, we expect from \cite{kw1} that
\beq\eqlabel{DVk1}
\kappa_1(\beta)=-\frac{1}{24}(\beta+\beta^{-1})\,.
\eq
The other two $k_{2,3}(\beta)$ can be fixed by considering the special moduli $s_1=-s_2=s(\delta)$ that is related to the $SU(2)$ Seiberg-Witten theory \cite{DGKV02}.  By matching the refined 1-loop amplitude at these special moduli to the $SU(2)$ partition function in the $\Omega$-background, we conclude 
\beq\eqlabel{DVk2k3}
\kappa_2(\beta)=-\frac{1}{2}\,,\,\,\,\,\,\kappa_3(\beta)=\frac{1}{12}(6-\beta-\beta^{-1})\,.
\eq
Comparing with the explicit perturbative matrix model expansion shows that \req{DV1loop} with \req{DVk1} and \req{DVk2k3} indeed reproduces the refinement of the cubic results, and hence we have verified that the refined B-model is indeed dual to the $\beta$-ensemble, at least at the 1-loop level.

For later comparison let us give the following explicit expansions of $\Fcal^{(1)}(s_1,s_2;\beta)$, obtained using \req{DV1loop} and \req{DVk1} and \req{DVk2k3} 
\beq\eqlabel{DVF1SWslice}
\lim_{\beta\rightarrow 0}\beta\,\partial^2_s\Fcal^{(1)}(s,-s;\beta)=\frac{1}{12}s^{-2}+153+\frac{46810}{3}s+1217160\, s^2+\dots\,,
\eq
\beq\eqlabel{DVF1S1}
\lim_{\beta\rightarrow 0}\beta\,\partial^2_{s_1}\Fcal^{(1)}(s_1,s_2;\beta)=\frac{1}{24}s_1^{-2}+\frac{97}{3}+\frac{4004}{3}s_1-\frac{6467}{3}s_2-\frac{299731}{2}s_1s_2+\dots\,,\\
\eq
where we have set for simplicity $\delta=g=1$. \footnote{Notice that, as explained in the previous sections, from the quantum mechanics point of view it is more natural to consider a double expansion in $g_s^2= -\ep_1 \ep_2$ and $\hbar = \ep_1+\ep_2$, while from the point of view of the holomorphic anomaly equation it is more natural to expand in $g_s$ and treat $\beta=-\ep_1/\ep_2$ as a parameter. Also notice that the parameter $\beta$ always enters in the symmetric combination $\beta+\beta^{-1}$ in the latter formalism. In order to compare the results obtained from these two viewpoints, simply notice that $g_s^2 (\beta+\beta^{-1}) = \hbar^2 + 2 g_s^2$. So the coefficients $\lim_{\beta\rightarrow 0}\beta\,\Fcal^{(1)}$ should really be compared with the $g_s^0\hbar^2$ term in the $g_s,\hbar$ expansion when taking the NS limit. More generally, the term $\lim_{\beta\rightarrow 0}\beta^g\,\Fcal^{(g)}$ contributes to the term $g_s^0 \hbar^{2g}$ in the $g_s$,$\hbar$-expansion.}

\paragraph{Seiberg-Witten slice}
Let us consider first the simplified case with $s=s_1=-s_2$ (the Seiberg-Witten slice). This restriction of parameters has been discussed in detail in \cite{DGKV02}. The one-form takes the simpler form
\beq\notag
\lambda=dx~\sqrt{(x^2-a^2)(x^2-b^2)}\,,
\eq
and the periods can be easily expressed through complete elliptic integrals of first and second kind, \ie,
\beq\eqlabel{DVSWperiods}
\begin{split}
\Pi_{A}^{(0)}&=\frac{1}{2\pi \ii}\int_a^b \lambda=\frac{a}{6\pi}\Big((a^2+b^2)E(k_1)-2b^2K(k_1)\Big)\,, \\
\Pi_B^{(0)}&=2\int_{-b}^b\lambda=\frac{2a}{3}\Big((a^2+b^2)E(k_2)-(a^2-b^2)K(k_2)\Big)\,,
\end{split}
\eq
with $k_1^2=1-\frac{b^2}{a^2}$ and $k_2^2=\frac{b^2}{a^2}$. The relation between the parameters $a,b$ and the original parameters $\delta,\Lambda$ reads
\beq\notag
a=-\frac{1}{2}\sqrt{\delta^2+8\Lambda^2}\,,\,\,\,\,\,b=-\frac{1}{2}\sqrt{\delta^2-8\Lambda^2}\,.
\eq
Inverting $\Pi_{A}^{(0)}(\Lambda)$, plugging it into $\Pi_B^{(0)}(\Lambda)$ and taking the $\Pi_{A}^{(0)}$ derivative yields the gauge kinetic function of the cubic matrix model under specialization $s_1=-s_2=s$ \cite{DGKV02}. 

Let us now apply the approach of section \ref{QsurfacesNS}, using the technique introduced in subsection \ref{The Gaussian}. It is convenient to absorb the energy $E$ into the potential $V(x)$ such that $V(x)=(x^2-a^2)(x^2-b^2)$ and $E=0$. Hence, at order $\hbar^2$ we have to integrate (see \req{partialS})
\beq\eqlabel{DVSWh2int}
\int dx~\frac{V''(x)}{48((x^2-a^2)(x^2-b^2))^{3/2}}\,.
\eq
It is convenient to define a differential operator $\Dcal^{(2)}$ acting on 
\beq\notag
\frac{1}{2\pi\ii}\int_a^b dx\frac{1}{\sqrt{(x^2-a^2)(x^2-b^2)}}=-\frac{1}{2\pi a}K(k_1)\,,
\eq
and
\beq\notag
2\int_{-b}^b dx\frac{1}{\sqrt{(x^2-a^2)(x^2-b^2)}}=\frac{2}{a}K(k_2)\,,
\eq
to obtain \req{DVSWh2int}. It is easy to see that 
\beq\notag
\Dcal^{(2)}=-\frac{1}{6ab(a^2-b^2)}\left(a(a^2-5b^2)\partial_{b}-b(b^2-5a^2)\partial_a \right)\,,
\eq
does the job. Hence, \req{DVSWh2int} evaluates to
\beq\notag
\begin{split}
S^{(2)}&=-\frac{1}{12\pi ab^2(a^2-b^2)}\left((a^4-10a^2b^2+b^4)E(k_1)+4b^2(a^2+b^2)K(k_1)\right)\,,\\
\Pi^{(2)}&=-\frac{1}{3 ab^2(a^2-b^2)}\left((a^4-10a^2b^2+b^4)E(k_2)-(a^4-6a^2b^2+5b^4)K(k_2)\right)\,.
\end{split}
\eq
Proceeding for $s:=\Pi_A^{(0)}+\hbar^2 \Pi_A^{(2)}$ and $\Pi_B=\Pi_B^{(0)}+\hbar^2 \Pi_B^{(2)}$ as sketched for the pure classical part above, yields the first order quantum correction to the gauge kinetic function. The result precisely matches the expectation \req{DVF1SWslice}.

\paragraph{General case}
Let us check that the agreement also holds for the general cases where $s_1$ and $s_2$ are independent. The starting point is as in the previous section the one-form expressed through the roots given in \req{DVwroot}. We have to integrate
\beq\eqlabel{wh2dvcubic}
\int dx \frac{V''(x)}{48((x-x_1)(x-x_2)(x-x_3)(x-x_4))^{3/2}}\,,
\eq
where now $V(x)=(x-x_1)(x-x_2)(x-x_3)(x-x_4)$. As above, this can be achieved via defining an operator $\Dcal^{(2)}$ acting on the classical periods. In order to construct such an operator, note that it is sufficient to find operators $\Dcal_I$ with $I=0,1,2$ satisfying
\beq\notag
\Dcal_{I}V(x)=x^I\,,
\eq
since $V''(x)$ is a polynomial of degree $2$. The operator $\Dcal^{(2)}$ is then a combination of the $\Dcal_I$. Such a basis of operators for the given potential can be inferred to be
\beq\notag
\begin{split}
\Dcal_0&=\frac{1}{(x_1-x_2)(x_3-x_4)} \left(\frac{-(x_2-x_3)\partial_{x_1}+(x_1-x_3)\partial_{x_2}-(x_1-x_2)\partial_{x_3}}{(x_1-x_3)(x_2-x_3)}\right.\\
&+\left.\frac{(x_2-x_4)\partial_{x_1}-(x_1-x_4)\partial_{x_2}+(x_1-x_2)\partial_{x_4}}{(x_1-x_4)(x_2-x_4)}\right),
\end{split}
\eq
\beq\notag
\begin{split}
\Dcal_1&=\frac{1}{(x_4-x_3)}\left(x_4\frac{(x_2-x_4)(\partial_{x_2}-\partial_{x_1})-(x_1-x_2)(\partial_{x_2}-\partial_{x_4})}{(x_1-x_2)(x_2-x_4)(x_1-x_4)}\right.\\
&+\left. x^3\frac{(x_2-x_3)\partial_{x_1}-(x_1-x_3)\partial_{x_2}+(x_1-x_2)\partial_{x_3}}{(x_1-x_2)(x_1-x_3)(x_2-x_3)}\right)\,,
\end{split}
\eq
\beq\notag
\Dcal_2=\frac{1}{\prod_{I<J}^4(x_I-x_J)}\left(\sum_{I=1}^4\prod_{\substack{J<K\\J,K\neq I}}^4(x_J-x_K) x_I^2\partial_{x_I}\right)\,.
\eq
The combined operator reads
\beq\notag
\Dcal^{(2)}=\frac{1}{3}\left(c_2  \Dcal_2+ c_1\Dcal_1+c_0\Dcal_0  \right)\Dcal_0\,,
\eq
with
\beq\notag
\begin{split}
c_0&=2(x_2x_3+(x_2+x_3)x_4+x_1(x_2+x_3+x_4))\,,\\
c_1&=-6(x_1+x_2+x_3+x_4)\,,\\
c_2&=12\,.
\end{split}
\eq
Applying $\Dcal^{(2)}$ to the classical periods yields the first quantum corrections. Inverting $s_I:=(1+\hbar^2\Dcal^{(2)})\Pi_{A_I}^{(0)}$ yields the quantum correction to the mirror map at order $\hbar^2$, \ie, 
\beq\notag
\begin{split}
z_1(s_1,s_2)|_{\hbar^2}&=2+62s_1-78s_2-4808s_1s_2+\dots\,,\\
z_2(s_1,s_2)|_{\hbar^2}&=2+78s_1-62s_2-4808s_1s_2+\dots\,.
\end{split}
\eq
Plugging these into $\Pi_1=(1+\hbar^2\Dcal^{(2)})\Pi_1^{(0)}$ and taking the $s_1$ derivative yields the expected expansion \req{DVF1S1}.

\subsection{Local $\P^1\times \P^1$}
\begin{figure}[t]
\psfrag{a}[cc][][0.7]{$\beta_1$}
\psfrag{b}[cc][][0.7]{$\alpha_1$}
\psfrag{c}[cc][][0.7]{$\alpha_2$}
\psfrag{d}[cc][][0.7]{$\alpha_3$}
\psfrag{e}[cc][][0.7]{$\alpha_4$}
\psfrag{T}[cc][][0.7]{$z_1,z_2\ll 1$}
\psfrag{f}[cc][][0.7]{$\t x=-1$}
\psfrag{g}[cc][][0.7]{$\t x=0$}
\psfrag{h}[cc][][0.7]{$\t x=\infty$}

\begin{center}
\includegraphics[scale=0.5]{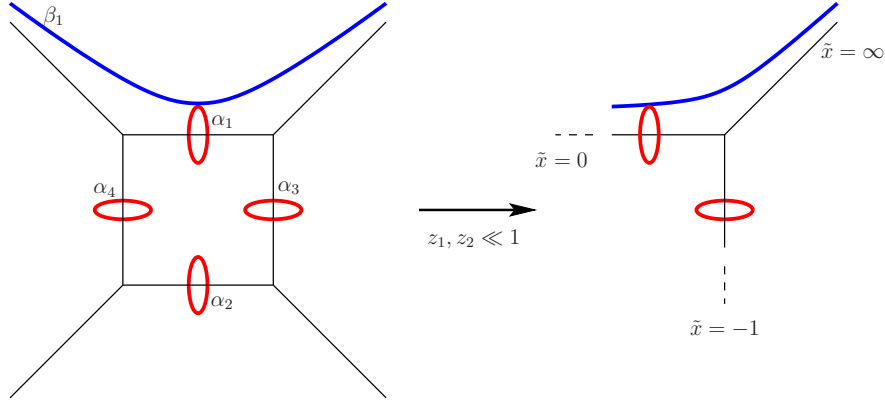}
\caption{The skeleton of the B-model geometry of local $\P^1\times\P^1$. A-period contours are drawn in red while B-period contours in blue.}
\label{F0fig}
\end{center}
\end{figure}

Let us consider non-trivial toric Calabi-Yau geometries of genus 1. For example, the mirror curve for local $\P^1\times\P^1$ can be parameterized as
\beq\eqlabel{P1P1curve}
H(p,x)=-1+e^x+e^p+z_1e^{-x}+z_2e^{-p}=0\,.
\eq
The skeleton of the corresponding Riemmanian surface is illustrated in figure \ref{F0fig} together with the integration contours for the classical periods \cite{akmv}.

Inspired by the matrix model results of section \ref{QsurfacesNS}, we consider the brane wave-fucntion satisfying the difference equation
\beq\eqlabel{F0diffeq}
(-1+e^x+z_1 e^{-x})\Psi(x)+\Psi(x+\hbar)+z_2\Psi(x- \hbar)=0\,,
\eq
obtained by canonically quantizing the curve \req{P1P1curve}. In the limit $x\rightarrow \infty$, we can view \req{F0diffeq} approximately as a difference equation with constant coefficients, whose solution takes the form
\beq\eqlabel{inftywavefunction}
\Psi(x)= \pi_1(x)e^{\frac{a_1}{\hbar} x}+\pi_2(x) e^{\frac{a_2}{\hbar}  x}\,,
\eq
with some functions $\pi_I(x)$ periodic under $x\to x+\hbar$ and some constants $a_I$.
In particular, for $a_1>a_2$ one can show that \cite{Norlund}
\beq\eqlabel{a1defi}
e^{a_1}=\lim_{x\rightarrow\infty} \frac{\Psi(x+\hbar)}{\Psi(x)}\,.
\eq
However, comparing \req{inftywavefunction} with the general form of the brane wave-function near $x\rightarrow\infty$ which can be inferred from \cite{akv01}, one infers that actually
\beq\notag
a_1=\frac{1}{2}\left(\Pi_{A_I}-\log z_I\right)\quad,\quad I=1,2\;,
\eq
where $\Pi_{A_I}$ denotes the fully quantum corrected A-periods. Explicit calculation of $a_1$ gives
\beq\notag
\begin{split}
a_1=&-(z_1+z_2)-\left(4+\frac{1}{q}+q\right)z_1z_2-\frac{3}{2}(z_1^2+z_2^2)\\
&-\left(16+\frac{1}{q^2}+q^2+\frac{6}{q}+6q\right)(z_1z_2^2+z_1^2z_2)-\frac{10}{3}(z_1^3+z_2^3)+\dots\,
\end{split}
\eq
as the first few orders in small $z_{1,2}$ expansion, where $q:=e^{\hbar}$, and we have kept only the finite terms. Expansion in $\hbar$ yields $\Pi_{A_I}=\sum_{n=0}^\infty \Pi_{A_I}^{(n)}\,\hbar^n$ with $\Pi_{A_I}^{(0)}$ the classical period and the first few quantum corrections 
\beq\eqlabel{P1P1quantA}
\begin{split}
\Pi_{A_I}^{(1)}&=0\,,\\
\Pi_{A_I}^{(2)}&=-2z_1 z_2-20(z_1^2z_2+z_1z_2^2)- 420 z_1^2z_2^2+\dots\,.\\
\Pi_{A_I}^{(3)}&=0\,.\\
&\vdots
\end{split}
\eq
Inverting $\Pi_{A_I}$ gives the mirror maps including quantum corrections thereof. We obtain
\beq\eqlabel{F0mmap}
\begin{split}
z_1(Q_1,Q_2)|_{\hbar^1}&=z_1(Q_1,Q_2)|_{\hbar^3}=0\,,\\
z_1(Q_1,Q_2)|_{\hbar^2}&=-2Q_1^2Q_2-4Q_1^2Q_2^2+\dots\,.
\end{split}
\eq 
with $Q_i= e^{-t_i}$ and $t_i$ are the K\"ahler parameters of the geometry.
One can also solve the difference equation order by order in $\hbar$ using a WKB Ansatz. For the leading (classical) piece we obtain 
\beq\notag
S'_0=p(x)=\log\left(- \frac{-1+e^x+z_1e^{-x}\pm e^{-x}\sqrt{(-e^x+e^{2x}+z_1)^2-4e^{2x }z_2}}{2}\right)\,,
\eq
where the two different signs correspond to the two solutions $\Psi_\pm(x)$. The higher order terms $S'_n$ can be easily obtained, though they are too lengthy to be displayed here. It is convenient to use the coordinate $\tilde x = e^x$ instead of $x$. Let us fix a branch, say $\Psi_+(\t x)$ and let us expand $S_n'$ for small $z_i$. Effectively, this means that we restrict to a local $\C^3$ patch of the geometry, as also shown in figure \ref{F0fig}. We observe that in this limit the expansion of $S'_n$ has poles at $\t x=0$ and $\t x=-1$. The poles correspond to the two `internal' punctures of the local $\C^3$ patch. Taking for instance the residue at $\t x=0$ yields the instanton part of the classical A-period, \ie,
\beq\eqlabel{P1P1periodA}
\frac{1}{2}\left(\Pi_{A_I}^{(0)}-\log z_I\right)= \oint_{\alpha_1} \frac{d\t x}{\t x}\, S_0'=-(z_1+z_2)-6z_1z_2-\frac{3}{2}(z_1^2+z_2^2)-30(z_1^2z_2+z_1z^2_2)+\dots\,.
\eq
Integrating the higher order $S_n'$ we can indeed reproduce our previous results \req{P1P1quantA}, as explicit calculation shows. \footnote{Depending on the actual parameterization of the curve, one might have to take the correct combination of `small' periods $\alpha_i$, \ie,  
$\Pi^{(n)}_{A_I}=\left(\oint_{\alpha_1}-\oint_{\alpha_2}\right) \frac{d \t u}{\t u}\, S_n'\,$ \cite{akv01}, to ensure that the odd sector in $\hbar$ drops out. The contour integral around $\alpha_2$ can be obtained by performing a $SL(2,\mathbb Z)$ transformation of the curve such that in the limit $z_1,z_2\ll 1$ we end up in the corresponding local patch.}

Let us now consider the B-period. We observe that we can recover the instanton part of the classical B-period in the local patch via the integral
\beq\notag
\int^\Lambda_{\delta}  \frac{d\t x}{\t x}\, S'_0= -z_2-\frac{1}{2}z_1^2-\frac{11}{4}z_2^2-5z_1z_2-\frac{77}{2}z_1z_2^2-\frac{47}{2}z_1^2z_2-\frac{1377}{4}z_1^2z_2^2+\dots\,
\eq 
up to a constant, and where we have taken $\delta\to0, \Lambda\to \infty$. Note that this integral diverges as $\delta\rightarrow 0$ and we kept only the finite terms.
Hence,
\beq\notag
\int^\Lambda_{\delta} \frac{d\t u}{\t u}\, S'_0 = \frac{1}{2}\left(\Pi_{B_1}^{(0)}-\frac{1}{2}\Pi_{A}^{(0)}\log(z_1)-\frac{1}{4}\log(z_1)^2\right)\,.
\eq
Similarly, we obtain for the order $\hbar^2$ 
\beq\notag
\Pi_{B_1}^{(2)}=-\frac{1}{6}z_2-\frac{8}{3}z_1z_2-\frac{1}{2}z_2^2-\frac{62}{3}z_1^2z_2-\frac{107}{3}z_1z_2^2-564z_1^2z_2^2+\dots\,.
\eq
Plugging the mirror map \req{F0mmap} into $\Pi_{B_1}=\Pi_{B_1}^{(0)}+\Pi_{B_1}^{(2)}\hbar^2 $ yields at order $\hbar^2$
\beq\notag
\Pi_{B_1}(Q_1,Q_2)|_{\hbar^2}=c -\frac{Q_2}{6}-\frac{7}{3}Q_1Q_2-\frac{Q_2^2}{6}-\frac{17}{2}Q_1^2Q_2-\frac{551}{3}Q_1^2Q_2^2+\dots\,,
\eq
where $c$ denotes some constant.

Let us compare this result to what we expect. In the B-model, it is easy to see that the refined 1-loop amplitude 
\beq\notag
\Fcal^{(1)}(\beta)=\frac{1}{2}\log(\det g)-\frac{\beta+\beta^{-1}}{24}\log(\Delta)-\frac{15-(\beta+\beta^{-1})}{24}\log(z_1z_2)\,,
\eq
with $g_{IJ}=\lim_{\hbar \to0}\partial_{Q_J} z_I|$ and $\Delta=1-8(z_1+z_2)+16(z_1-z_2)^2$ reproduces the refined vertex results of \cite{ikv}. Taking the NS limit, we conclude
\beq\notag
\Pi_{B_1}(Q_1,Q_2)|_{\hbar^2}= Q_2\partial_{Q_2} \lim_{\beta\rightarrow 0}\beta\, \Fcal^{(1)}\,,
\eq
up to the constant part of $\frac{1}{24}$\footnote{The reason why we do not get  the constant part in our calculation is a technical subtlety due to our method of B-cycle integration rather than a conceptual issue.}. 

Let us also check the order $\hbar^4$. For simplicity, we will focus on the special moduli slice of the moduli space with $z=z_1=z_2$. The mirror map at this order comes out to be
\beq\notag
z(Q)|_{\hbar^4}=-\frac{Q^3}{6}-\frac{14}{3}Q^4+\dots\,.
\eq
We obtain for the B-period
\beq\notag
\Pi^{(4)}_B=-2z-\frac{33}{2}z^2-\frac{1280}{9}z^3+\dots \,,
\eq
Hence,
\beq\notag
\Pi^{(4)}_B(Q)=\frac{Q}{360}-\frac{19}{60}Q^2-\frac{199}{20}Q^3+\dots\,.
\eq
Comparison with the refined vertex results of \cite{ikv} shows that this indeed matches the $\beta\rightarrow 0$ limit to the corresponding order under the identification $Q:=Q_1=Q_2$. 

\subsection{Local $\P^2$}
\begin{figure}[t]
\psfrag{e}[cc][][0.7]{$\alpha_2$}
\psfrag{c}[cc][][0.7]{$\alpha_1$}
\psfrag{b}[cc][][0.7]{$\alpha_3$}
\psfrag{d}[cc][][0.7]{$\beta$}
\psfrag{f}[cc][][0.7]{$\beta_1$}
\psfrag{j}[cc][][0.7]{$\beta_2$}
\psfrag{g}[cc][][0.7]{$\t x=0$}
\psfrag{h}[cc][][0.7]{$\t x=-1$}
\psfrag{T}[cc][][0.7]{$z \ll 1$}
\psfrag{1}[cc][][0.7]{$i)$}
\psfrag{2}[cc][][0.7]{$ii)$}
\begin{center}
\includegraphics[scale=0.5]{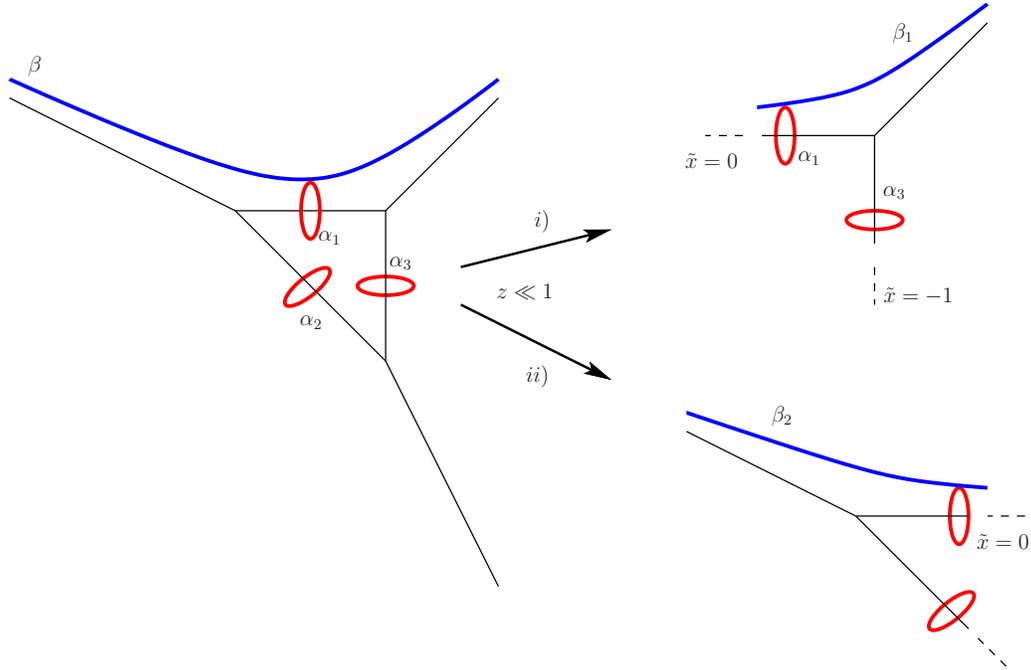}
\caption{The skeleton of the B-model geometry of local $\P^2$. A-period contours are drawn in red while B-period contours in blue. Case {\bf (i)} shows the effective geometry of the parametrization \req{P2curve1} under the limit $z\ll 1$, while {\bf (ii)} of \req{P2curve2}.}
\label{P2fig}
\end{center}
\end{figure}

Let us consider local $\P^2$ as our final example. We parameterize the mirror curve as
\beq\eqlabel{P2curve1}
H(p,x)=-1+e^x+e^p+ze^{\hbar/2}e^{-x-p}=0\,.
\eq 
The corresponding geometry is sketched in figure \ref{P2fig}.
A canonical quantization yields the difference equation
\beq\notag
(-1+e^x)\Psi(x)+\Psi(x+\hbar)+ze^{-x}e^{\hbar/2}\Psi(x-\hbar)=0\,.
\eq
As in the previous example, we solve for $a_1$ defined in \req{a1defi} to extract the quantum A-period given by $a_1=\frac{1}{3}\left(\Pi_A-\log(z)\right)$. We obtain
\beq\notag
\begin{split}
a_1&=-\frac{1+q}{\sqrt{q}}z-\left(6+\frac{1}{q^2}+\frac{7}{2q}+\frac{7}{2}q+q^2\right)z^2\\
&-\frac{3+9q+36q^2+88q^3+144q^4+144q^5+88q^6+36q^7+9q^8+3q^9}{3q^{9/2}}z^3+\dots\,.
\end{split}
\eq
Hence, via expansion in $\hbar$ we infer for $\Pi_A=\sum_{n=0}^\infty \Pi_A^{(n)}\,\hbar^n$ the first few corrections
\beq\eqlabel{P2qAperiods}
\begin{split}
\Pi^{(1)}_A&=0\,,\\
\Pi^{(2)}_A&=-\frac{3}{4}z-\frac{45}{2}z^2-630z^3-17325z^4+\dots\,,\\
\Pi^{(3)}_A&=0\,,\\
\Pi^{(4)}_A&=-\frac{z}{64}-\frac{39}{8}z^2-\frac{2961}{8}z^3-19635z^4+\dots\,.\\
&\vdots
\end{split}
\eq
Inverting $\Pi_A(z)$ gives the quantum corrected mirror map. We obtain for the first few orders in $\hbar$ and in $Q$ expansion
\beq\eqlabel{P2mmap}
\begin{split}
z(Q)|_{\hbar^1}&=z(Q)|_{\hbar^3}=0\,,\\
z(Q)|_{\hbar^2}&=-\frac{3}{4}Q^2-9Q^3-171Q^4+\dots\,,\\
z(Q)|_{\hbar^4}&=-\frac{1}{64}Q^2-\frac{15}{4}Q^3-\frac{3429}{16}Q^4+\dots\,.\\
\end{split}
\eq

Let us now solve the difference equation via the WKB Ansatz given by \req{WKB1} and \req{WKB2}. The leading term of the solution reads
\beq\notag
S_0'(x)=\log\left(\frac{1-e^x\pm e^{-\frac{x}{2}}\sqrt{e^{x}(1-e^x)^2-4z}}{2}\right)\,.
\eq
Higher order solutions $S'_n$ can be obtained easily, but are too length to be explicitly displayed here. Similar as in the previous example, expansion for $z\ll 1$ effectively zooms into a $\C^3$ patch of the geometry and the classical A-period can be read of from a contour integral around one of the singular points (\cf, figure \ref{P2fig}), \ie,
\beq\notag
\frac{1}{3}\left(\Pi^{(0)}_A-\log(z)\right)=\oint_{\alpha_1}\frac{d\t x}{\t x}\,S_0'\,.
\eq
Similarly, the higher order $\Pi_A^{(n)}$ can be obtained via integrating $S_n'$ and we can indeed reproduce \req{P2qAperiods}. Let us turn to the B-period. The integration contour is indicated in blue in figure \ref{P2fig}. We split the integration contour into two parts 
\beq\notag
\beta=\beta_1+\beta_2\,.
\eq
Taking the limit $z\ll 1$ of the parametrization of the curve given in \req{P2curve1} restricts to a $\C^3$ patch which includes $\beta_1$ (\cf, figure \ref{P2fig}).

Explicit evaluation shows that for integrating along $\beta_1$
\beq\notag
\int_\delta^\Lambda\frac{d\t x}{\t x}\,S_0'=\frac{1}{3}\left(\Pi_B^{(0)}-\frac{1}{3}\log(z)\Pi^{(0)}_A -\frac{1}{6}\log(z)^2\right)\,,
\eq
where $\Lambda\rightarrow\infty$, $\delta\rightarrow 0$ and we kept only finite terms. Let us calculate
\beq\notag
\Pi^{(n)}_{\beta_1}=\int_\delta^\Lambda \frac{d\t x}{\t x}\,S_n'\,.
\eq
For example, for even $n$ we obtain the first quantum corrections
\beq\notag
\begin{split}
\Pi^{(2)}_{\beta_1}&=c-\frac{11}{24}z-\frac{87}{8}z^2-\frac{3349}{12}z^3-\frac{176005}{24}z^4+\dots\,,\\
\Pi^{(4)}_{\beta_1}&=-\frac{127}{5760}z-\frac{577}{160}z^2-\frac{71081}{320}z^3-\frac{758401}{72}z^4+\dots \,,
\end{split}
\eq
with $c$ some constant. In contrast to local $\P^1\times\P^1$ discussed in the previous section, we do not have anymore a $\mathbb Z_2$ symmetry of the integration contour $\beta$ and hence we can no longer expect that the integration along $\beta_1$ and $\beta_2$ yields up to signs the same result. Thus, we have to explicitly calculate the integrals along $\beta_2$.  For that, we have to change the parametrization of the curve \req{P2curve1} such that we end up in the limit $z\ll 1$ in the patch which includes $\beta_2$. It is not hard to see that the transformation $x\rightarrow -x, p\rightarrow -p$ combined with a change of complex structure achieves this. We obtain the curve
\beq\eqlabel{P2curve2}
-1+ze^{-x}+e^{-p}+e^{ \hbar/2}e^{x+p}=0\,,
\eq
which we quantize and solved via a WKB Ansatz as above. We obtain
\beq\notag
\int_\delta^\Lambda\frac{d\t x}{\t x}\,S_0'=-\frac{2}{3}\left(\Pi_B^{(0)}-\frac{1}{3}\log(z)\Pi^{(0)}_A -\frac{1}{6}\log(z)^2\right)\,.
\eq
Hence, adding up the integrations along $\beta_1$ and $\beta_2$ yield indeed the classical B-period. Integrating along $\beta_2$ for some higher $S_n'$ yields
\beq\notag
\begin{split}
\Pi^{(2)}_{\beta_2}&=c-\frac{5}{12}z-\frac{57}{4}z^2-\frac{2509}{6}z^3-\frac{141355 }{12}z^4+\dots\,,\\
\Pi^{(4)}_{\beta_2}&=\frac{67}{2880}z-\frac{317}{80}z^2-\frac{45401}{160}-\frac{532285}{36}z^4+\dots\,.\\
\end{split}
\eq
Adding up $\Pi^{(n)}_B=\Pi^{(n)}_{\beta_1}+\Pi^{(n)}_{\beta_2}$ and inserting the quantum corrected mirror map \req{P2mmap}, we finally obtain (up to the constant part of $\Pi_B^{(2)}$)
\beq\eqlabel{P2results}
\begin{split}
\Pi^{(2)}_B&=c-\frac{7}{8}Q-\frac{129}{8}Q^2-\frac{589}{2}Q^3+\dots\,,\\
\Pi^{(4)}_B&=\frac{29}{640}Q-\frac{207}{32}Q^2-\frac{55341}{160}Q^3+\dots\,.\\
\end{split}
\eq
Let us compare this result to what we know. From the results of \cite{kw1} one expects that the refined B-model 1-loop amplitude is given by
\beq\notag
\Fcal^{(1)}(\beta)=\frac{1}{2}\log\left(\tau\right)+\frac{-16+\beta+\beta^{-1}}{24}\log\left(z\right)-\frac{\beta+\beta^{-1}}{24}\log\left(1-27z\right)\,,
\eq
with $\tau:=Q\partial_Q z(Q)|_{\hbar^0}$ and one can easily infer that this indeed matches the corresponding refined vertex results. Using the usual holomorphic anomaly equations, it is straightforward to calculate in the B-model higher genus amplitudes $\Fcal^{(g)}(\beta)$, which can again be matched with refined vertex results \cite{kwunp} (this also has been confirmed in \cite{hk10}). Comparing with \req{P2results} shows that
\beq\notag
\begin{split}
\Pi^{(2)}_B&= Q\partial_Q \lim_{\beta\rightarrow 0}\beta\, \Fcal^{(1)}(\beta)\,,\\
\Pi_B^{(4)}&= Q\partial_Q \lim_{\beta\rightarrow 0}\beta^2\, \Fcal^{(2)}(\beta)\,.
\end{split}
\eq

\section{Brane Wave-Functions and Liouville Amplitudes}\label{lio}

In section \ref{MMtimedepS} we have seen that the brane partition functions of the $\beta$-ensemble matrix models satisfy an operator equation which takes the form of the conformal Ward identity equation.
 When specializing to the Penner type logarithmic potential, it is well-known that the matrix model computes the Liouville conformal blocks with background charge
$\tilde Q= b+b^{-1}$, where $$ b^2 =-\beta= \frac{\ep_1}{\ep_2}\;.  $$

In this case, the $\ep_1$- and $\ep_2$-brane are exactly given by the two types of degenerate vertex operators, denoted by $V_{-\frac{1}{2b}}$ and $V_{-\frac{b}{2}}$ respectively, and the  differential equation \req{differential_conformal_form} becomes nothing but the usual BPZ equation \cite{Belavin:1984vu} for conformal blocks with degenerate field insertions
$$
\Bigl(b^{\pm 2} \del_x^2 - \sum_a \frac{1}{ x-z_a}\del_{z_a} - \sum_a \frac{\Delta_{\alpha_a}}{(x-z_a)^2} \Bigr) Z_\alpha(x) =0 \;,
$$
where $\Delta_\alpha= \alpha (\tilde Q-\alpha)$ is the conformal dimension of the operator with momentum $\alpha$.
We have also argued that, provided we know the answer for the open string partition function, this knowledge can help us to deduce the closed string partition function, especially in the NS limit. We will see below how this idea is realized in this case.

The key observation is that the $\epsilon_1$-brane partition function \req{bmm} is always a polynomial of degree $N$ in the brane location $x$.
To be explicit, let us choose the logarithmic matrix model potential $-\frac{2}{\epsilon_2}W(x) = 2 b \sum_{a=0}^{n-2}  \alpha_a\log(x-z_a)$ corresponding to the $n$-point function $\langle V_{\alpha_{n-1}}(\infty)\prod_{a=0}^{n-2} V_{\alpha_a}(z_a)\rangle$, where the vertex inserted at infinity carries momentum given by the conservation rule $\sum_{a=0}^{n-1} \alpha_a =\tilde Q-N b$. 
Suppose we are interested in the corresponding $n$-point conformal block computed by an appropriate choice of contour following \cite{Cheng2010}, we can probe the 
$n$-point function by inserting an $\ep_1$-brane, or $V_{-\frac{1}{2b}}(x)$ in the Liouville language, and arrive at the brane partition function
\bea\notag 
Z_1(x) &=& \prod_{a<b} (z_a - z_b)^{2\alpha_a \alpha _b}\prod_a(x- z_a)^{ \alpha_a/b}\\\label{wf}&&\times \int d^Nz \; \prod_{1\leq i<j\leq N}(z_i - z_j)^{-2b^2} \;\prod_{i, a}(z_i - z_a)^{2 \alpha_a b} \;\prod_i(z_i -x)\;
\eea
which is explicitly a polynomial in $x$.

As alluded above, this $\ep_1$-brane has the property that we can move it close to any other point $z_a$ of insertions without introducing any non-analyticity, and this gives us various relations among the $n$-point function with momenta shifted in units of $-\frac{1}{2b}$. In this way we can obtain essential information about the $n$-point function (the closed partition function) from the knowledge of the $n+1$-point function (the brane partition function). 

In more details, consider the $n$-point function 
$$
Z_{closed}( \alpha) =\prod_{a<b} (z_a - z_b)^{2\alpha_a \alpha _b}\ \int d^Nz \; \prod_{1\leq i<j\leq N}(z_i - z_j)^{-2b^2} 
\;\prod_{i=1}^N\prod_{a=0}^{n-2}(z_i - z_a)^{2 \alpha_a b}\;,
$$
it is easy to see that it satisfies the following relations to $Z_1(x)$ given in (\ref{wf}):
\be\label{tpt}
Z_1(z_a) = \lim_{x\rightarrow z_a} Z_1(x) = Z_N(\alpha_a-\tfrac{1}{2b})\;,
\ee
and
$$
Z_{closed}( \alpha) =\prod_{a<b} (z_a - z_b)^{2\alpha_a \alpha _b}\ \int d^Nz \; \prod_{1\leq i<j\leq N}(z_i - z_j)^{-2b^2} 
\;\prod_{i, a}(z_i - z_a)^{2 \alpha_a b}\;.
$$

%
%
%
%
Thus we find a set of difference equations for $\log Z_{closed}$, one for each non-degenerate insertion point, which depends on the open partition function $Z_1(x)$ but not on its normalization
$$
Z_{closed}(\alpha_a-\tfrac{1}{2b})/Z_{closed}(\alpha_a) = Z_1(z_a)/Z_1(\infty)\;.
$$
On the left hand side, we keep everything fixed in both the numerator and the denominator, apart from the momentum of the insertion point $Z_1(x)$ approaches.\footnote{From the point of view of Liouville conformal blocks, $Z_1(\infty)$ is related to $\langle V_{\alpha_{n-1}-1/2b}(\infty)\prod_{a=0}^{n-2} V_{\alpha_a}(z_a)\rangle$ where the momentum of the insertion at infinity has been shifted by $-1/2b$.} From these, at least in principle, we can determine (up to normalization) $Z_{closed}$ from the  knowledge of $Z_1(x)$. 
We will now illustrate this procedure with the of $3+1$-point function example. In this case the potential $W(x)$ has only one critical point and hence the corresponding spectral curve has genus zero. Recall that from the general discussion in section \ref{MMexamples}, the time dependence of the Schr\"odinger equation is trivial and one does not need to take the NS limit. 

In this case, the open partition function
\be\notag
Z_1(x) = x^{\Delta_{\alpha_0-{1}/{2b}}-\Delta_{\alpha_0}-\Delta_{{1}/{2b}}} (1-x)^{\Delta_{\alpha_1-{1}/{2b}}-\Delta_{\alpha_1}-\Delta_{{1}/{2b}}}\Psi_N(x) \, \ee
with
\be\notag
\Psi_N(x) = \int d^Nz \; \prod_{1\leq i<j\leq N}(z_i - z_j)^{-2b^2} \;\prod_{i =1}^N (z_i)^{2 \alpha_0b} \;(z_i-1)^{2\alpha_0b}  \;\prod_i(z_i -x)
\ee
satisfies the differential equation
\bea\notag
\!\!\left(\!-{b^{2}} \frac{d^2}{d x^2}+\big(\frac{1}{x}+\frac{1}{x-1}\big) \frac{d}{d x}-\frac{\Delta_{\alpha_0}}{x^2}-\frac{\Delta_{\alpha_1}}{(x-1)^2} +\frac{\Delta_{\alpha_0}+\Delta_{\alpha_1}+\Delta_{-1/2b}-\Delta_{\alpha_2}}{x(x-1)}\right)Z_1(x)=0
\eea
with 
\be\label{alpha2}
\alpha_0+\alpha_1+\alpha_2-1/2b= \tilde Q-Nb \;. 
\ee
This is solved by 
$$
\Psi_N(x) = c\, F(A,B;C;x)
$$
with 
$$
A= -N\;,\; B = N-1+ \tfrac{2}{b} (\alpha_0+\alpha_1 -  \tfrac{1}{b})\;,\; C = \tfrac{2}{b}  (\alpha_0 - \tfrac{1}{2b})\;,
$$
and $F(A,B;C;x)$ is the hypergeometric series 
$$F(A,B;C;x) = \sum_{n=0}^{\infty} {(A)_n (B)_n\over (C)_n} {x^{n}\over n!}\;,$$
with $(A)_n = A(A+1)\ldots (A+n-1)$.
The normalization constant $c$ remains to be fixed.   
Since $A=-N$ is a negative integer in our case, the solution $\Psi_N(x)$ is a finite polynomial in $x$ of degree $N$, as we argued earlier. From this, we can find the recurrence equations satisfied by the chiral half of the Liouville 3-point function
$$
Z_N(\alpha_0,\alpha_1)= \int d^Nz \; \prod_{1\leq i<j\leq N}(z_i - z_j)^{-2b^2} \;\prod_{i =1}^N (z_i)^{2 \alpha_0b} \;(z_i-1)^{2\alpha_1b} \;.
$$
Using \req{tpt} and colliding $x$ with $0$, $1$ and $\infty$ and using analyticity of $\Psi_N(x)$, it immediately follows that   %
$$\frac{Z_N(\alpha_0+1/2b, \alpha_1)}{ Z_N(\alpha_0, \alpha_1) }=\frac{\psi_N(0)}{ \psi_N(\infty)}$$
$$\frac{Z_N(\alpha_0, \alpha_1+1/2b)}{ Z_N(\alpha_0, \alpha_1) }=\frac{\psi_N(1)}{ \psi_N(\infty)}\;.
$$

Using the special values
\bee\notag
F(A,B;C;x=1) &=& \frac{\Gamma(C)\Gamma(C-A-B)}{\Gamma(C-A)\Gamma(C-B)}\\  \lim_{x\to \infty} x^{-N} F(A=-N,B;C;x) &=& (-1)^N\frac{(B)_N}{(C)_N} \;,
\eee
the above recursive relation reads
\bee\notag
\frac{Z_N(\alpha_0-\tfrac{1}{2b},\alpha_1)}{Z_N(\alpha_0,\alpha_1)} &=& \prod_{j=0}^{N-1} \left(\frac{2b\alpha_0-1+j b^2}{2b\alpha_2-1+jb^2}\right)\\
\frac{Z_N(\alpha_0,\alpha_1-\tfrac{1}{2b})}{Z_N(\alpha_0,\alpha_1)} &=& \prod_{j=0}^{N-1} \left(\frac{-2b\alpha_1+1-j b^2}{2b\alpha_2-1+jb^2}\right)\;,
\eee
where $\alpha_2$ is given by $\alpha_{0,1}$ as in (\ref{alpha2}).

From this we conclude that the chiral three-point function must satisfy
$$
{Z_N(\alpha_0,\alpha_1)} \sim \prod_{j=0}^{N-1} \frac{\Gamma(1-2b\alpha_1-jb^2)}{\Gamma(2b\alpha_0+j b^2)\Gamma(2b\alpha_2+j b^2)}\;.
$$
Indeed, after fixing the value $Z_N({0,0})$ as the initial condition, we then obtain the expression
$$
Z_N(\alpha_0,\alpha_1) = \prod_{j=0}^{N-1} \frac{\Gamma(-(1+j)b^2)\Gamma(1-2b\alpha_1-jb^2)}{\Gamma(-b^2)\Gamma(2b\alpha_0+j b^2)\Gamma(2b\alpha_2+j b^2)}\;,
$$
which indeed gives the chiral half of the Liouville three-point function \cite{Schiappa:2009cc,Cheng2010}.

\section{Conclusion}
\label{conclusion}
In this work we have discussed several aspects of refined topological strings, emphasizing the role of
the branes. In particular, we have argued that brane partition functions are quantum mechanical wave-functions satisfying multi-time dependent Schr\"odinger-like equations, with times given by non-renormalizable moduli. The derivation was done using a matrix model realization of the refined topological string on specific geometries. However, we expect this to hold in more generalities, though in general the precise identification of the times is not immediately transparent yet.

In the NS limit, the time dependence drops out and we end up with a time-independent Schr\"odinger equation, making contact with the earlier results of \cite{adkmv,EM08,Chekhov:2009mm}. The refined partition function in this limit can be recovered from the brane wave-function via
considering the monodromies around the cycles of the local geometry, which we have also checked explicitly in several non-trivial examples.
Using this fact, we have explained the observation of Nekrasov and Shatashvili in connecting integrable
systems to gauge theory partition functions:  The integrable system arises in the study of the world-sheet amplitudes
and their target interpretation can be phrased in terms of gauge theory partition functions, thus explaining the NS result.

There are various extensions of the present work one can consider.  The simplest one, which should be straightforward, is to generalize
the discussions in this paper from a single matrix model to Toda-like matrix models.  This should in principle allow
us to get arbitrary B-model geometries with arbitrary analytic $H(x,p)$, where the powers of $p,x$ are bounded.

A more important extension of this work involves a deeper understanding of the brane partition function
away from the NS limit.  As already mentioned above, even away from this limit we expect an interesting wave-function
which is now time-dependent.  Uncovering the meaning of this wave-function from the perspective of integrable systems could
be very interesting.  Also, finding a way to compute via the time-dependent wave-function the complete refined topological string partition function, as we have done for the NS limit via the time-independent wave-function, would be clearly important.  We are currently elaborating on these ideas and plan to report on them elsewhere.

\begin{acknowledgments}
We like to thank S. Lee, A. Okounkov, S. Shakirov, S. Shatashvili and L. Takhtajan for useful discussions.
We would also like to thank the Simons Center for Geometry and Physics, where this work was initiated
in the 8th Simons Workshop in Mathematics and Physics. 
M.A. and D.K. would like to thank Harvard for hospitality during part of this work. D.K. likes to thank ASC Munich, CERN-TH and the MF Oberwolfach for hospitality during the final stage of this project. The research of M.A. and D.K. was supported by the Berkeley Center for Theoretical Physics, by the National Science Foundation (award number 0855653), by the Institute for the Physics and Mathematics of the Universe, and by the US Department of Energy under Contract DE-AC02-05CH11231. The research of M.C. was supported in part by NSF grant DMS-0854971. The research of D.K. was  supported by a Simons fellowship. The research of C.V. was supported in part by NSF grant PHY-0244821. 
\end{acknowledgments}

\end{document}